\documentclass[twocolumn,color,superscriptaddress,psfig,showpacs,amsmath,amssymb,floatfix,fleqn]{revtex4-1}

\usepackage{amsmath}
\usepackage{amsfonts}
\usepackage{amssymb}
\usepackage{amsthm}
\usepackage{mathtools}

\usepackage{epsf}
\usepackage{graphicx}
\usepackage{epstopdf}

\usepackage{natbib}
\setcitestyle{square,numbers}

\usepackage{color}

\newcommand{\LCSO}{LiCuSbO$_4$}
\newcommand{\LSCO}{LiSbCuO$_4$}
\newcommand{\Cu}{Cu$^{2+}$}

\newcommand{\Licuv}{LiCuVO$_{4}$}
\newcommand{\slrr}{$T_1^{-1}$}
\newcommand{\li}{$^{7}$Li}
\newcommand{\slm}[1]{\textcolor{magenta}{#1}}
\definecolor{auburn}{rgb}{0.43, 0.21, 0.1}

\definecolor{dgreen}{rgb}{0,0.4,0.3}

\begin{document}

\title{Evidence for a magnetic field-induced unconventional nematic
state in the frustrated and anisotropic spin-chain cuprate \LCSO}

\author{H.-J.\ Grafe}
\affiliation{Leibniz Institute for Solid State and Materials Research
IFW-Dresden, D-01171 Dresden, Germany}

\author{S.~Nishimoto}
\affiliation{Leibniz Institute for Solid State and Materials Research IFW-Dresden, D-01171 Dresden, Germany}
\affiliation{Institute for Theoretical Physics, Technical University Dresden, D-01069 Dresden, Germany}

\author{M.\ Iakovleva}
\affiliation{Leibniz Institute for Solid State and Materials Research IFW-Dresden, D-01171 Dresden, Germany}
\affiliation{Zavoisky Physical-Technical Institute of the Russian Academy of Sciences, 420029 Kazan, Russia}

\author{E.\ Vavilova}
\affiliation{Leibniz Institute for Solid State and Materials Research IFW-Dresden, D-01171 Dresden, Germany}
\affiliation{Zavoisky Physical-Technical Institute of the Russian Academy of Sciences, 420029 Kazan, Russia}

\author{L.\ Spillecke}
\affiliation{Leibniz Institute for Solid State and Materials Research IFW-Dresden, D-01171 Dresden, Germany}
\affiliation{Institute for Solid State
Physics, Technical University Dresden, D-01069 Dresden, Germany}

\author{A.\ Alfonsov}
\affiliation{Leibniz Institute for Solid State and Materials Research IFW-Dresden, D-01171 Dresden, Germany}

\author{M.-I.\ Sturza}
\affiliation{Leibniz Institute for Solid State and Materials Research IFW-Dresden, D-01171 Dresden, Germany}

\author{S.\ Wurmehl}
\affiliation{Leibniz Institute for Solid State and Materials Research IFW-Dresden, D-01171 Dresden, Germany}

\author{H.\ Nojiri}
\affiliation{Institute of Materials Research, Tohoku University, 980-8577, Sendai, Japan}

\author{H.~Rosner}
\affiliation{Max-Planck-Institute for Chemical Physics of Solids, Dresden, Germany}

\author{J.\ Richter}
\affiliation{Universit\"at Magdeburg, Institut f\"ur Theoretische Physik, Germany}

\author{U.K.\ R{\"o}{\ss}ler}
\affiliation{Leibniz Institute for Solid State and Materials Research IFW-Dresden, D-01171 Dresden, Germany}

\author{S.-L.~Drechsler}
\affiliation{Leibniz Institute for Solid State and Materials Research IFW-Dresden, D-01171 Dresden, Germany}

\author{V.\ Kataev}
\email[Corresponding author:]{ v.kataev@ifw-dresden.de}
\affiliation{Leibniz Institute for Solid State and Materials Research IFW-Dresden, D-01171 Dresden, Germany}

\author{B.\ B\"{u}chner}
\affiliation{Leibniz Institute for Solid State and Materials Research IFW-Dresden, D-01171 Dresden, Germany}
\affiliation{Institute for Solid State Physics, Technical University Dresden, D-01069 Dresden, Germany}

\date{\today}

\begin{abstract}

Modern theories of quantum magnetism predict exotic multipolar states in weakly interacting strongly frustrated spin-1/2 Heisenberg chains with
ferromagnetic nearest neighbor (NN) inchain exchange in high magnetic fields. Experimentally these states remained elusive so far. Here we report the
evidence for a long-sought magnetic field-induced nematic state arising above a field of $\sim 13$\,T in the edge-sharing chain cuprate
\LSCO\,$\equiv$\,\LCSO. This interpretation is based on the observation of a field induced spin-gap in the measurements of the $^7$Li NMR spin
relaxation rate \slrr\ as well as a contrasting field-dependent power-law behavior of \slrr\ vs.\ $T$ and is further supported by static
magnetization and ESR data. An underlying theoretical microscopic approach favoring a nematic scenario is based essentially on the NN XYZ exchange
anisotropy within a model for frustrated spin-1/2 chains. It is investigated by the DMRG technique. The employed exchange parameters are justified
qualitatively by electronic structure calculations for \LCSO.
\end{abstract}


\maketitle
%

Strong electronic correlations in solids may give rise to novel ground states of matter such as electronic liquid crystal phases in correlated metals
or spin liquid states in insulating quantum magnets \cite{Fradkin10,Sachdev08}. Theory predicts that in the latter systems conventional long-range
magnetic dipolar order can be suppressed down to $T=0$ due to frustration of magnetic interactions and/or quantum fluctuations (for a review see,
e.g., \cite{Balents10}). Though individual spins remain non-ordered in the spin liquid, higher rank magnetic multipoles (quadrupoles, octupoles etc.)
can order under favorable conditions \cite{Penc11}.  Such an exotic multipolar (MP) order does not break the time-reversal symmetry and is often
referred to as a "hidden order" since it is difficult to detect experimentally by most of the available techniques sensitive to magnetic dipole
moments, only. However, the spin rotational symmetry is broken in this hidden phase which is therefore also called a spin-nematic state in the
simplest quadropolar case, in analogy with the nematic order in liquid crystals, where the translational crystalline order is absent but the
rotational symmetry is broken.

A prominent example of a spin liquid is a single Heisenberg spin-1/2 chain with the nearest neighbor (NN) antiferromagnetic (AFM) interaction whose
ground state is described by the so-called gapless Tomonaga-Luttinger (TL) spin liquid \cite{Mikeska04}. Further complexity can be brought in the
problem by including an AFM next-nearest neighbor (NNN) interaction $J_2$. Irrespective of the sign of the NN coupling $J_1$, AFM $J_2$ causes spin
frustration and may yield different phases depending on the frustration ratio $\alpha = |J_2/J_1|$ \cite{Majumdar69,Haldane82,Chubukov91}. In
particular, pioneering theoretical works \cite{Chubukov91,Kecke07,Kuzian07,Vekua07,Hikihara08,Laeuchli09,Sudan09,Ueda09,Zhitomirsky10} devoted mainly
to one-dimensional (1D) isotropic frustrated $J_1$(FM)\,-\,$J_2$(AFM) chain models have predicted unusual field-induced MP states near the saturation
field $H_{\rm sat}$ above which at $T=0$ all spins are aligned by an external magnetic field $H$. These states form a TL-liquid of multiple $p$-bound
states of magnons corresponding to nematic, triatic, quartic MP phases ($p= 2, 3, 4, ...$ ). For interacting chains within 2D or 3D arrangements, the
MP phases strongly compete with collinear longitudinal $H$-dependent incommensurate spin density wave (SDW$_p$) phases predominant at lower fields
\cite{Starykh14}. (Here the index $p$ indicates the neighboring MP state.) Albeit MP
 phases might coexist with {\it non-collinear } strongly
fluctuating dipolar states \cite{Smerald16}.

\begin{figure}[b!]
\includegraphics[width=0.84\columnwidth]{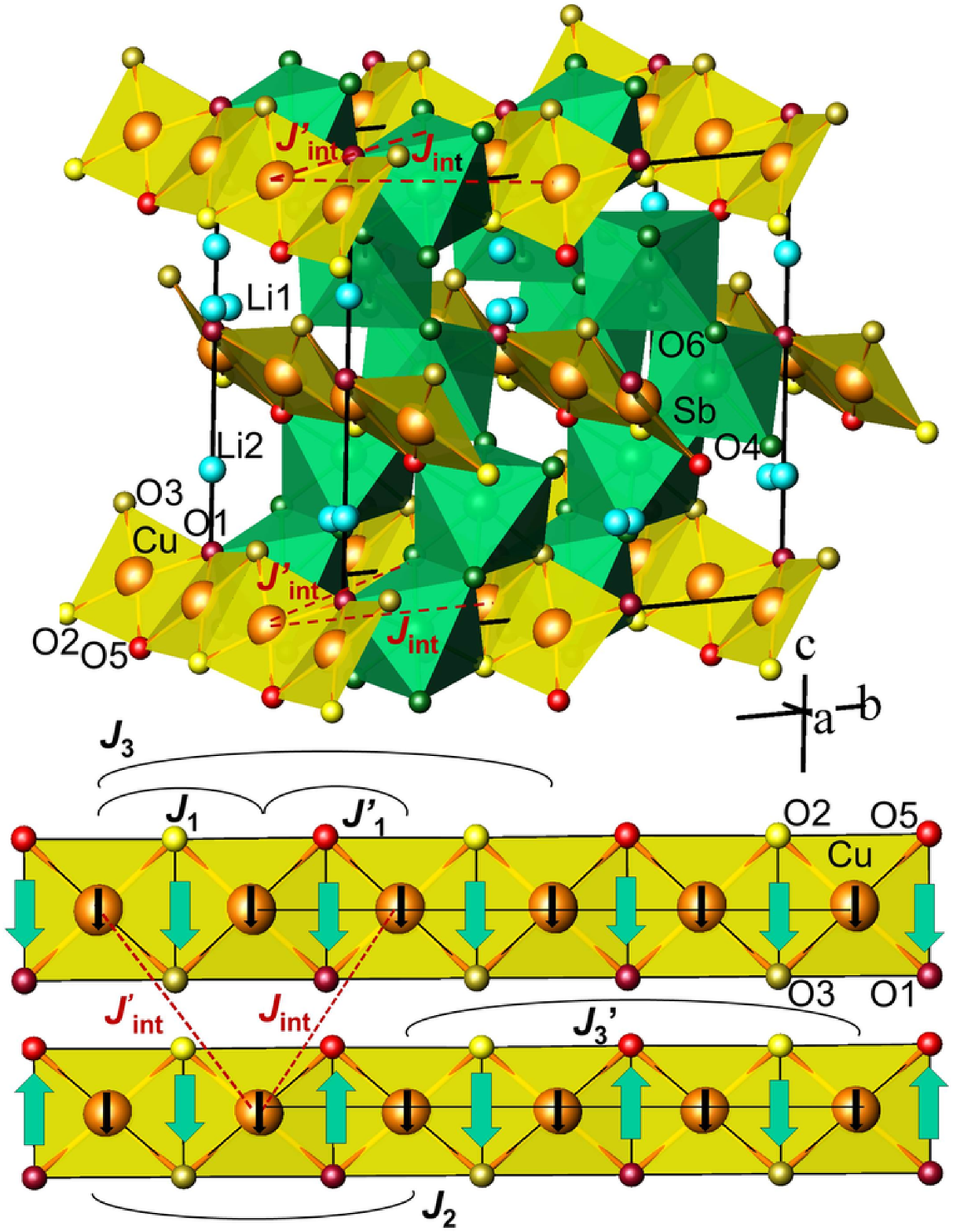}
\caption{Top: crystallographic structure
 of \LSCO\,$\equiv$\,\LCSO\ \cite{remark}.
 \Cu\ ions (orange) are bonded
covalently to four {\it nonequivalent} O$^{2-}$ ligands and form buckled non-planar CuO$_4$ plaquettes. Edge-shared CuO$_4$ plaquettes form non-ideal
alternatingly tilted stacks (along the $c$-axis) of CuO$_{2}$ {\it chains} (colored yellow and brown, respectively) running along the $a$-axis. The
chains are interconnected with SbO$_6$ octahedra shown in blue-green. The Li$^{+}$ ions (bright blue balls) occupy two positions. The split Li1
position is shown with overlapping balls. Bottom: Schematic view of two neighboring individual CuO$_2$ chains within the $ab$-plane ignoring their
tilting and buckling (cf. Top). The relevant intrachain exchange paths are indicated by black arcs. The four nonequivalent O$^{2-}$ ions within a
CuO$_4$-plaquette give rise to different left and right Cu-Cu bonds along the $a$-axis causing this way an {\it alternated} $(J_1,J'_1)$-$J_2$ spin
chain with small nonequivalent third neighbor couplings $(J_3,J'_3)$. Black arrows  denote the magnetically active spins and bright blue arrows the
DM vectors that are confined to the ($bc$)-plane $\perp$ to the chain axis whereas their mutual orientation can be arbitrary. For illustration the
two limiting configurations, uniform and staggered are depicted that have been studied theoretically in single-chain approximation
(Fig.~\ref{mag_nematic}). Dark dashed line: the frustrated weak interchain couplings $J_{int}$ and $J'_{int} \sim 1$~K$\ll J_2 < \mid J_1\mid,\mid
J'_1\mid $ in the basal $(ab)$-plane (see also Top).  } \label{structure}
\end{figure}

Low-D spin networks can be found in 3D transition metal (TM) oxides.
 A specific geometry of the chemical bonds can yield chains
of TM ions magnetically coupled mainly via oxygen ligands in one direction.  A sizable and still increasing number of frustrated CuO$_2$ spin-1/2
chain compounds is currently available. Nevertheless, the very existence of an MP state is not yet proved. Also the possibility of a coexistence of
an MP state (and even of an ordered phase) with some other dipolar magnetic phases or corresponding strong fluctuations of them remains unclear up to
now. The quest for MP phases focussed in the last years mainly on two compounds, namely \Licuv\ \cite{Nawa13,Buettgen14,Prozorova15} and
PbCuSO$_4$(OH)$_2$ (linarite) \cite{ Willenberg16}. Both materials exhibit at ambient fields a 3D-non-collinear spiral type dipolar order at low $T$
due to residual interchain couplings. Various field-induced collinear SDW$_p$ phases were detected but the theoretically proposed neighboring MP
phases at higher fields remain elusive. In particular, in the case of \Licuv\ it was concluded that the spin-nematic phase, if it exists at all,
could be established only in a very narrow field range of 1\,T below $\mu_0H_{\rm sat}\approx 41.4$\,T \cite{Buettgen14,commentMourigal12}.
 Indeed, from the theoretical side it has become
clear that the interchain coupling that causes dipolar magnetic order can
easily destroy fragile MP phases whereas easy-axis exchange anisotropy may
stabilize them \cite{Nishimoto15,Onishi15a}. In this context the recent
synthesis and the first physical study of a novel member of the  frustrated
CuO$_2$-chain family, the strongly frustrated $J_1$(FM)\,-\,$J_2$(AFM)
quasi-1D compound \LCSO\ (Fig.~\ref{structure}), is noteworthy
\cite{Dutton12,remark}. It exhibits short-range incommensurate spin
correlations below $T \sim 9$\,K but, in contrast to the spiral spin-chain
compounds \Licuv\ and LiCuZrO$_4$ \cite{Drechsler07,Vavilova09}, do not show
long-range magnetic order at $H=0$ down to $ T \sim 0.1$\,K.
 This points to a very weak or specific interchain coupling in
\LCSO. Moreover, a sizeable exchange anisotropy was estimated here, too.

\begin{figure}[!h]
\includegraphics[width=0.9\columnwidth]{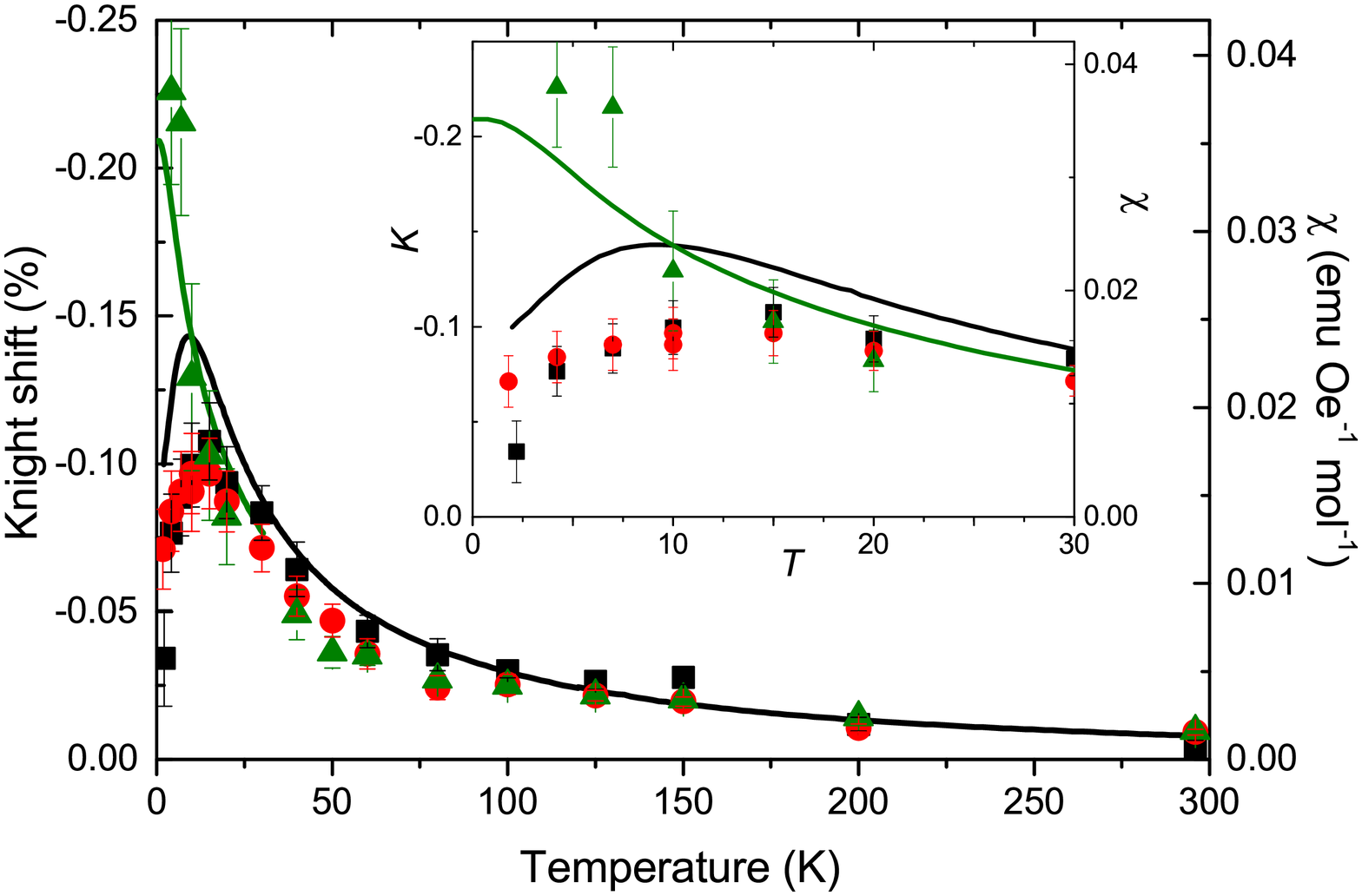}
\caption{Scaling of Knight shift ($K$(3T) $\blacksquare$, $K$(9T) \textcolor[rgb]{1,0,0}{$\bullet$}, and $K$(15T)
\textcolor[rgb]{0,0.58,0}{$\blacktriangle$}) and macroscopic susceptibility ($\chi$(3T) \textbf{---}, $\chi$(16T)
\textcolor[rgb]{0,0.58,0}{\textbf{---}}).
The susceptibility at 16 Tesla has been reproduced from Dutton \textit{et al.}~\cite{Dutton12}.}
\label{Knightsusce}
\end{figure}

Here we report the results of $^7$Li NMR relaxation rate \slrr\ measurements
in \LCSO\ in a broad field range of $3 - 16$\,T. A surprising
contrasting behavior of \slrr\ vs.\ $T$ is observed around a special crossover
field $H_{\rm c1} \approx 13$\,T within a rather narrow field range.
At $H < H_{\rm c1}$, \slrr$(T)$ exhibits a diverging power-law behavior
suggesting a static magnetic order at lower temperatures. In contrast, in
higher magnetic fields $H > H_{\rm c1}$, \slrr$(T)$ turns into an exponential
decrease indicative of the opening of a spin gap  above the narrow crossover
high-field region. Our analysis of the
static magnetization and ESR data rules out saturation of the magnetization
or Dzyaloshinskii-Moriya (DM) couplings as possible conventional reasons
for the opening of this spin gap. We argue that this gap should be considered
as one of the signatures of a distinctive but
nevertheless naturally "hidden" for a dipolar sensitive probe MP state.
In particular, we argue that such a peculiar behavior of \slrr\ is due to the
occurrence of a dipolar spin-liquid state confined to a certain field
range below $H_{\rm c1}$ which crosses over to the competing anisotropic
spin-nematic liquid state which is stabilized above $H_{\rm c1}$.
Our theoretical analysis employing the density matrix renormalization group (DMRG)
technique indeed reveals a broad stability region of an unconventional
spin-nematic state in \LCSO\ setting in above $\sim 13$\,T and extending to $
H \gtrsim 20$\,T. Altogether, our experimental observations and model
calculations provide strong arguments to
identify a long-sought nematic state in \LCSO\ and stress the importance
of anisotropic exchange for its  relevance.\\

\noindent{\bf Results}\\

\begin{figure}[t]
\includegraphics[width=0.8\columnwidth]{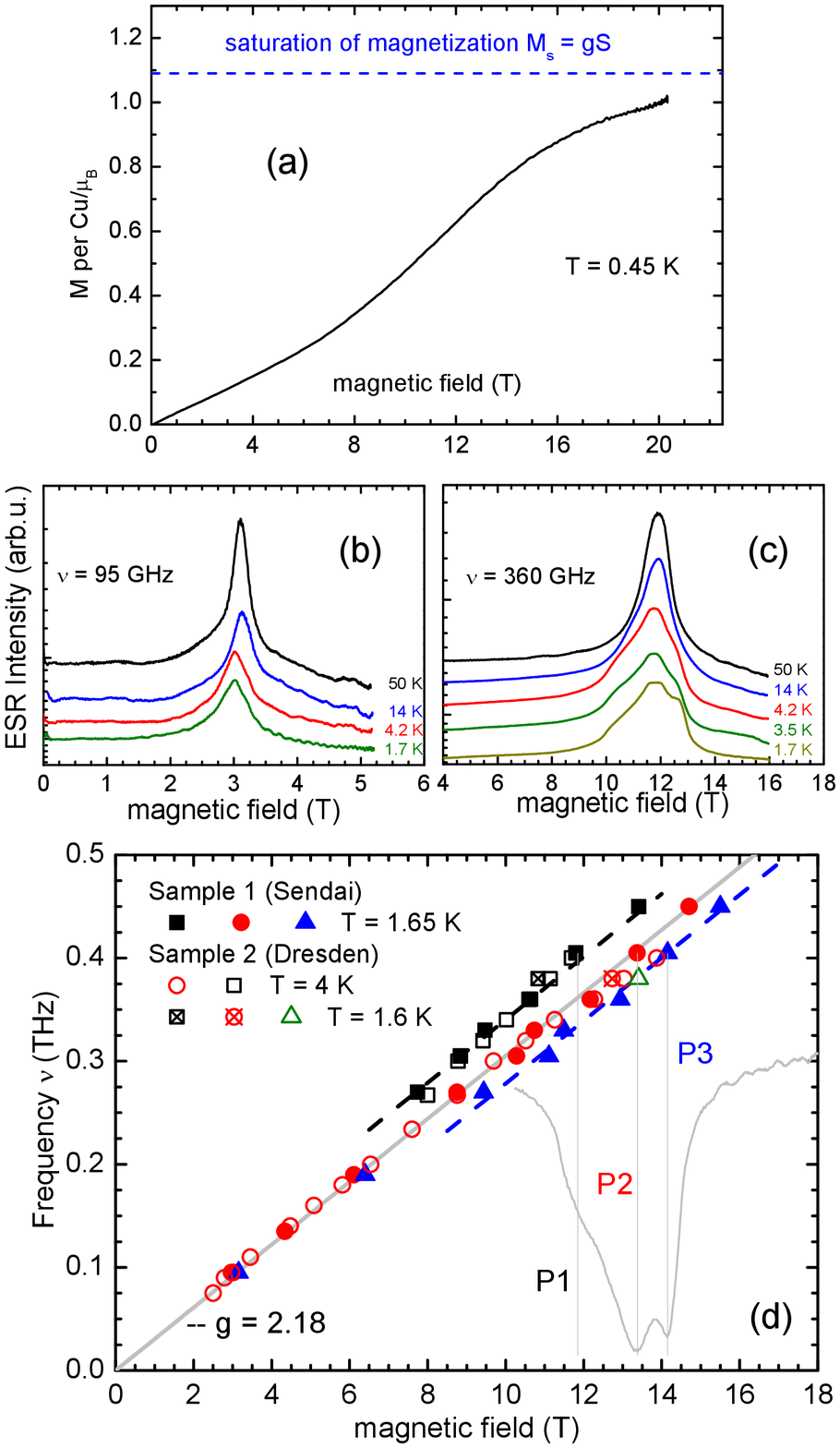}
\caption{Field dependence of the magnetization at $T = 0.45$\,K (a).
Selected ESR spectra at different temperatures at frequencies $\nu = 95$\,GHz
(b), and $\nu = 360$\,GHz (c), and a summary of the frequency vs.\ field
dependence of the ESR peaks at a low $T$ plotted together with a spectrum at
$\nu = 405$\,GHz (d). Note the satellite peaks in the spectra in panel (c)
that develop at $T < 20$\,K. In (d), symbols - experimental peak positions
P1, P2 and P3, solid and dashed lines are linear fits. The main peak P2
follows a linear gapless branch $\nu_2 = (\mu_{\rm B}/h)gH$ with $g = 2.18$,
whereas the satellite branches P1 and P3
$\nu_{1,3} = \Delta_{1,3} + (\mu_{\rm B}/h)gH$ exhibit offsets $\Delta_1 \approx 35$\,GHz and $\Delta_3
\approx -27$\,GHz. (see the text)
}
\label{FigESR}
\end{figure}

\noindent{\bf Magnetization measurements}. The $T$-dependence of the static magnetic susceptibility $\chi(T)= M/H$ of \LCSO\ measured at a field
$\mu_0H = 3$\,T is shown in Fig.~\ref{Knightsusce}. It appears to be in accord with the data of Ref.~\cite{Dutton12}. In particular, a characteristic
maximum of $\chi(T)$ is observed at $\sim 9$\,K. The dependence of the static magnetization $M$ on the magnetic field $H$ measured in pulsed fields
up to 20\,T at a low $T = 0.45$\,K is shown in Fig.~\ref{FigESR}(a). The $M(H)$ curve exhibits a characteristic $S$-shape. At fields below $\sim
6$\,T $M$ increases almost linearly and develops an upward curvature at higher fields as expected for a simple isotropic 1D-chain \cite{Griffits64}.
However, by approaching $\sim 12$\,T the $M(H)$ dependence weakens but surprisingly {\it no} saturation is observed up to the highest field of 20\,T.
Such a peculiar $M(H)$ dependence appears to be a remarkable feature of \LCSO\ as discussed in detail below \cite{commentMagn}.
\\

\begin{figure*}
\centering
\includegraphics[width=\textwidth]{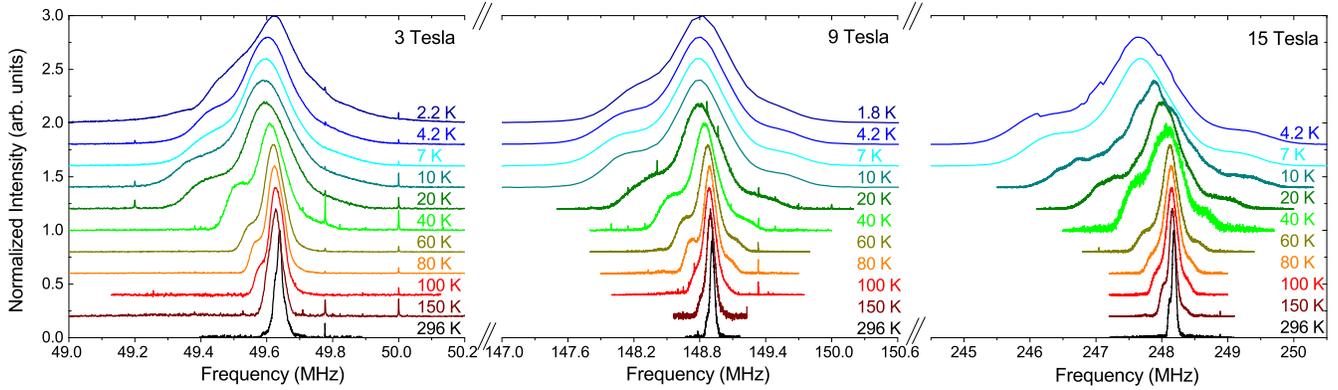}
\caption{NMR spectra at 3, 9 and 15 Tesla. Note that the width of the $x$-axis is scaled by a factor 9/3 and 15/3 for 9 and 15 Tesla with respect to
the $x$-axis of the 3~Tesla spectra. This way, this figure shows that the broadening is paramagnetic, and that there is no magnetic order down to
$\sim 2$ K in any field.} \label{NMRspectra}
\end{figure*}

{\bf ESR measurements}. High-field ESR spectra of \LCSO\ at all measured frequencies at temperatures $T > 20$\,K consist of a symmetrical single line
with the shape close to a Lorentzian [Figs.~\ref{FigESR}(b) and (c)]. Considering the polycrystalline form of our samples, this suggests that the
anisotropy of the $g$-factor, which otherwise produces a characteristic asymmetric pattern of the spectrum \cite{Poole96}, is smaller than the width
of the signal. The $g$-factor of 2.18 obtained from the slope of the
 $\nu(H)$ dependence $ g = (h/\mu_{\rm B})\nu/H $ [Fig.~\ref{FigESR}(d)] is a
typical powder-averaged value for a \Cu\ ion in a distorted octahedral ligand coordination \cite{Abragam70}. At $\nu = 95$\,GHz which corresponds to
the resonance field $\mu_0H_{\rm res} \approx 3.1$\,T the signal remains practically a single line down to the lowest temperature. At $T \lesssim
20$\,K it merely exhibits a rather small shift to smaller fields and slightly broadens, but shows no indication of an onset of a static magnetic
ordering. In contrast, at $\nu \geq 270$\,GHz and $\mu_0H_{\rm res} \geq 8.8$\,T the satellite peaks P1 and P3 begin to develop besides the central
peak P2 at $T \lesssim 20$\,K [Figs.~\ref{FigESR}(c) and (d)]. Their offset of $\sim 0.8 - 1$\,T from the main peak remains practically constant,
whereas the intensity increases with increasing $\nu$ (and $H$) and decreasing $T$. The frequency vs.\ field dependence of P1 and P3 approximately
follows the relation $\nu_{1,3} = \Delta_{1,3} + (\mu_{\rm B}/h)gH$ with $\Delta_1 \approx 35$\,GHz and $\Delta_3 \approx -27$\,GHz.

\noindent{\bf NMR measurements}. Frequency swept \li\ NMR spectra in a
field of 3, 9, and 15 Tesla are shown in Fig.~\ref{NMRspectra} (see methods).
The width of the spectra (square root of the second moment)
in 3\,T and at 296\,K is 26.5\,kHz, indicating that any quadrupolar broadening or
splitting must be significantly smaller than this value. Therefore, the
shape of the powder pattern is solely determined by the anisotropic dipolar
hyperfine coupling of the two different Li sites in LiCuSbO$_4$ times
the susceptibility in the paramagnetic state. The width of the spectra scales
perfectly linear with the magnetic field, indicating that the broadening
of the spectra is entirely of magnetic origin. The scaling holds also at low
temperatures, evidencing the absence of magnetic ordering in all fields down to $\sim 2$\,K.
The dipolar hyperfine coupling tensor has been
obtained by lattice sum calculation and is given in the Supplement.

The Knight shift, $K$, has been extracted
from the maximum of the spectra. According to the calculated dipolar hyperfine coupling tensor, both Li
sites contribute to the maximum with different elements of the tensor. $K$ is plotted together with the macroscopic susceptibility,
in Fig.~\ref{Knightsusce}. The scaling of the Knight shift with
 $\chi$, i.e.,
 $K = A_{hyp} \cdot \chi$, is good down to $\sim 15$\,K. Below 15\,K, the $K$ shows qualitatively the
same $T$-dependence as  but the error bars of $K$ become large at low temperatures due to the strong magnetic
broadening. However, the similar $T$-dependencies indicates that both, $K$ and $\chi$,
are dominated by the intrinsic
susceptibility and that impurity contributions are small.
\begin{figure}[b]
\includegraphics[width=0.9\columnwidth]{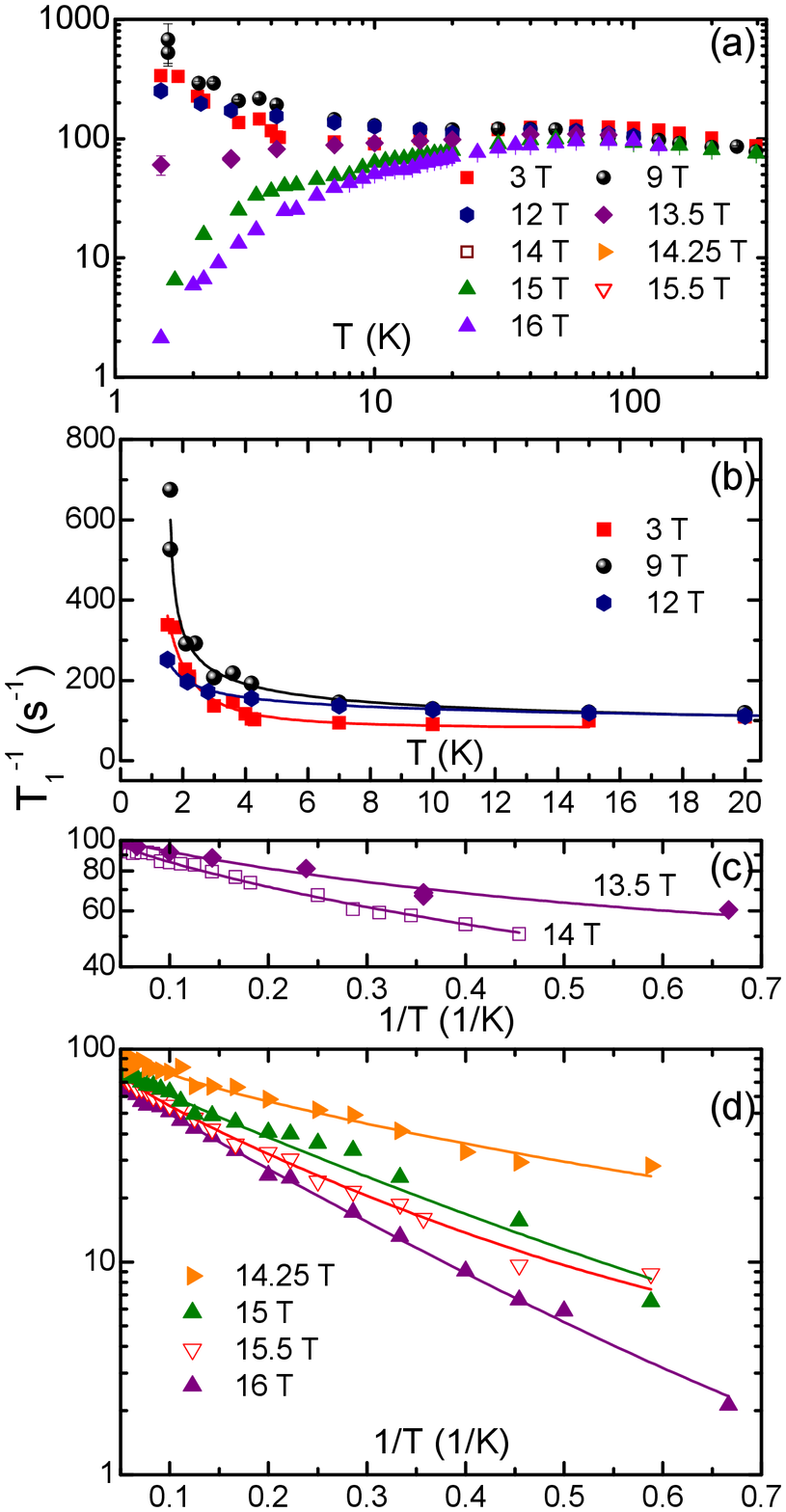}
\caption{(a) \slrr\ vs. temperature for different magnetic fields; (b)
The \slrr$(T)$ dependence at $T < 20$\,K for 3\,T, 9\,T and 12\,T; (c,d)
\slrr\ vs. inverse temperature $T^{-1}$ at $T < 20$\,K for fields $>13$\,T.
Solid lines in (b), (c) and (d) are model curves according to
Eq.~(\ref{mainfit}).} \label{T1}
\end{figure}

The results of the measurements of the \li\ nuclear spin-lattice relaxation rate \slrr\ are shown in Fig.~\ref{T1}. In Fig.~\ref{T1}(a) \slrr\ vs.
temperature is plotted on a log-log scale for six different fields from 3 to 16 Tesla. At high $T$, \slrr\ is almost constant and field-independent
as expected for a 1D spin system. However, at low $T$, below $\sim 30$ K, a dramatic field dependence develops, resulting in a difference of almost
three orders of magnitude between 3 and 16\,T at the lowest temperatures. For low fields below 12\,T (Fig.~\ref{T1}(b)), \slrr\ diverges at low
temperatures (for 3~T, \slrr\ increases below 7~K, for 9 and 12~T, \slrr\ increases already below 15 K). In contrast, for fields above 13.5~T, \slrr\
decreases almost exponentially at low $T$. This is highlighted in Figs.~\ref{T1}(c,d), where \slrr\ is plotted vs.\
the inverse temperature $T^{-1}$. \\
\begin{figure}
\includegraphics[width=0.8\columnwidth]{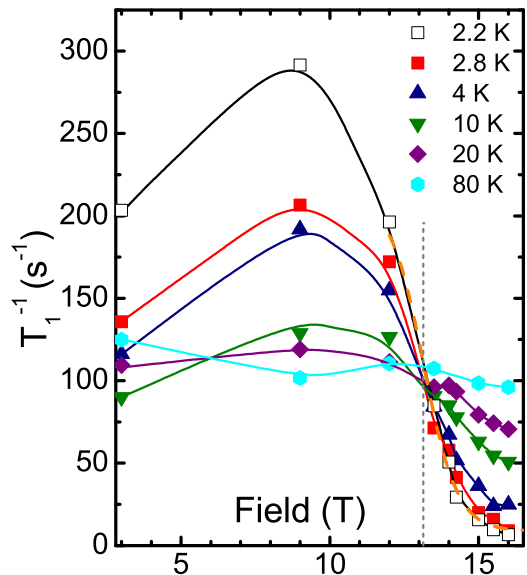}
\caption{The extracted spin-lattice relaxation rate $1/T_1$  vs.\ external magnetic field for selected temperatures. Full lines
are guides for the eye. The isosbestic point (IBP) near $\sim 13$\,T is indicated by the vertical dashed line. The dashed orange
curve is the fit to Eq.~(\ref{tanh}). (see the text) } \label{Figisos}
\end{figure}

\noindent{\bf Discussion}\\

\noindent{\bf Experiment.} The central experimental result of this work is an observation of distinct temperature and magnetic field dependences of
the \li\ NMR relaxation rate \slrr\ in the short-range ordered (SRO) state of \LCSO\ below $\sim 10$\,K, as shown in Fig.~\ref{T1}, which will be
discussed in the following.\\
\noindent\underline{\it NMR relaxation in magnetic solids}. In magnetic materials, the nuclear spin lattice relaxation rate, \slrr , is typically
caused by the transverse (i.e.\ $\perp$ to the nuclear spin quantization axis) components of the time-dependent fluctuating field exerted on the
nuclei by the electron spin system. It can be expressed in terms of the Fourier transforms $S^{z,\pm}(q,\omega)$ of the time-dependent longitudinal
and transverse spin-spin correlation functions $\langle S^{z}_{j}(t)S^{z}_{0}(0)\rangle$, $\langle S^{+}_{j}(t)S^{-}_{0}(0)\rangle$ and $\langle
S^{-}_{j}(t)S^{+}_{0}(0)\rangle$, respectively \cite{Moriya56}:
%
\begin{flalign}
\nonumber S^{z,\pm}(q,\omega) & =  \sum_j e^{-iqj}\int^\infty_{-\infty}dt e^{i \omega t}\langle S^{z,\pm}_{j}(t)S^{z,\mp}_{0}(0)\rangle_T , & \\
\nonumber T_1^{-1} & \propto  \frac{\gamma_n^2}{\gamma_e^2}  \lim_{\omega\rightarrow0}\sum_{q}|A_{\parallel}(q)|^2S^{z}(q,\omega) + & \\
 & |A_{\perp}(q)|^2[S^{\pm}(q,\omega) + S^{\mp}(q,\omega)] . & \label{T1eqn}
\end{flalign}
%
Here $q$ is the wave vector, $\langle \cdots \rangle_T$ means the thermal average, $\gamma_{e/n}$ are the gyromagnetic ratios of the electron and the
probed nucleus, and $A_{\parallel,\perp}(\vec{q})$ are the hyperfine form factors of the probed nucleus. The subscripts $\parallel$ and $\perp$
denote the hyperfine tensor components generating the transversal components of the fluctuating local field at the nuclear site due to the
longitudinal $\langle zz \rangle$ and transversal $\langle +- \rangle$ spin-spin correlations, respectively. For $\hbar\omega_{\rm NMR}\ll k_{\rm
B}T$, $S^{z,\pm}(q,\omega) \propto \chi''_{z,\pm}(q,\omega)k_{\rm B}T/\hbar\omega_{\rm NMR}$, where $\chi''_{z,\pm}(q,\omega)$ is the imaginary part
of the dynamical electron spin susceptibility, $\omega_{\rm NMR}$ is the NMR frequency, $k_{\rm B}$ and $\hbar$ are the Boltzmann's constant and the
reduced Planck constant, respectively. Thus at small NMR frequencies, $1/T_1 \propto \sum_q \chi ''(q,\omega \rightarrow 0)T$.

Generally, filtering effects may occur such that the hyperfine coupling is
peaked (or zero) for certain $q$-vectors. However, the crystal structure
of \LCSO\  indicates that both Li sites are coupled to several Cu sites
from different chains, suggesting a rather weak dependence of
$A_{\parallel,\perp}$ on $q$. Since the coupling between the \li\ nuclear
spin and the Cu spins is of dipolar nature both the longitudinal and the
transversal terms in Eq.~(\ref{T1eqn}) are expected to contribute to the
relaxation. Indeed, our estimates with the dipolar hyperfine model have
revealed comparable contributions from $\langle zz \rangle$ and $\langle  +- \rangle$ correlations
to the transversal field at both Li sites (see, Suppl.).\\

\noindent\underline{\it Field dependence of\ \slrr }. In weakly coupled
unfrustrated critical simple AFM Heisenberg chains in the paramagnetic state
far above the Ne\'{e}l ordering temperature $T_{\rm N} \ll T \ll J/k_{\rm B}$, the rate \slrr\ in general {\it continuously increases} with
decreasing $T$ and/or increasing the magnetic field $H$ up to the saturation field and tends to diverge by approaching $T_{\rm N}$. This is mainly
due to the growth of the $\langle S^{+}_{j}(t)S^{-}_{0}(0)\rangle$ correlation function with increasing $H$ and decreasing $T$ whereas $\langle
S^{z}_{j}(t)S^{z}_{0}(0)\rangle$ decays smoothly following a power law \cite{Chitra97,Sato09,Sato11}.

In \LCSO, however, \slrr\ shows a non-monotonous and even contrasting behavior with respect to temperature and magnetic field. At
relatively small fields the low-temperature region is determined by a more or less sharp increase of \slrr$(T)$ (Fig.~\ref{T1})
pointing to the vicinity of a critical magnetically ordered state at a lower temperature. Especially at $H = 9$\,T the increase
is substantially more pronounced than at lower fields such as for 3\,T indicating an increase of the ordering temperature of the
presumable magnetic phase. Such a behavior is not expected for an ordinary antiferromagnetic Ne\'{e}l state where $T_{\rm N}$ is
usually suppressed by an external magnetic field. In fact, the field region around 9\,T is also identified by the low-temperature
anomaly in the magnetic specific heat in Ref.~\cite{Dutton12}. It has been conjectured in that work to be a signature of an
unusual field-induced magnetic phase in \LCSO.

One further and even more striking feature, which can be easily recognized in Fig.~\ref{T1}, is the occurrence of a threshold
field of $\sim 13$\,T that separates the upturn behavior from a drastic suppression of \slrr\ vs.\ $T$. Indeed, plotting these
data points for fixed temperatures as a function of the field, yields a set of curves with a sharp general crossing point at
$H_{\rm c1} \approx 13$\,T, usually denoted in the literature as an {\it isosbestic} point (IBP) \cite{Greger13}
(Fig.~\ref{Figisos}). Since the nuclear \slrr\ is governed by fluctuating fields at the nuclear site produced by electron spins,
the IBP at 13\,T may be identified as the critical field that separates two regimes with {\it different} types of magnetic
fluctuations to be discussed in detail below. Indeed, at low-temperatures $T \leq 10 - 15$\,K which are of special interest here,
the observed IBP  coincides also with an inflection point (IFP). Using the generic linear field dependence $1/T_1 -const \propto
\left( H- H_{\rm c1} \right)$ in the vicinity of an IFP at $H_{\rm c1}$, it is tempting to generalize that linear behavior
further into the nonlinear region at higher fields employing a quasi-exponential expression that captures also the field region
slightly smaller than $H_{\rm c1}$:
\begin{eqnarray}
T_1^{-1}(H) = \frac{1}{T_1\left( H_{\rm  c1} \right)} \left[ 1 +0.92 \tanh \frac{H - H_{\rm  c1} }{1.12 A} \right] \ .
\label{tanh}
\end{eqnarray}
Here $1/T_1(H_{\rm c1}=13\ \mbox{T}) =110$~s$^{-1}$ and $A$ is a dimensional constant taken as 1\,T. As can be seen in
Fig.~\ref{Figisos}, Eq.(\ref{tanh}) describes the data in the considered field region for the lowest available temperature
$T=2.2$\,K quite well. This way we arrive at a
smooth transition across $H_{\rm c1}$ to a pronounced exponential-type behavior at high magnetic
fields and low temperature.\\

\begin{figure}
\includegraphics[clip,width=0.8\columnwidth]{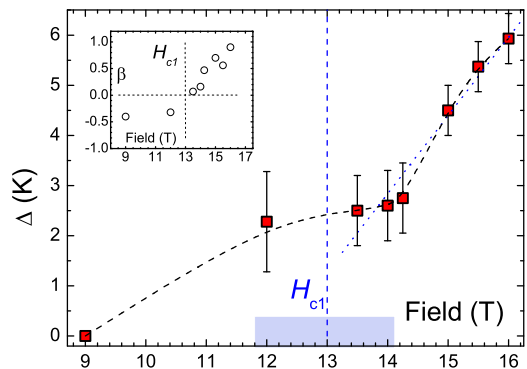}
\caption{Magnetic field dependence of the gap $\Delta$ (squares)
evaluated using Eq.~(\ref{mainfit}). Dashed line connecting the
data points is guide for the eye. Dotted line is a linear fit
revealing the slope $\Delta /H = 1.56$\,K/T; Inset: the same for
the exponent $\beta$ in Eq.~(\ref{mainfit}) (circles). Vertical
dashed lines denote the isosbestic point $H_{\rm c1} \approx
13$\,T. Shaded bar indicates a crossover region between the two
distinct regimes of the \li\ $T_1$ relaxation. (see the text)}
\label{fitgap}
\end{figure}

\noindent\underline{\it Temperature dependence of\ \slrr}. Importantly, as can be seen  in a logarithmic plot of \slrr\ vs. $T^{-1}$ in
Figs.~\ref{T1}(c,d), a similar predominantly exponential behavior develops for the strongest fields also in the $T$-dependence of \slrr\ suggesting
the opening of an energy gap for spin excitations. For a consistent analysis of the whole set of experimental \slrr$(T)$ curves we have used a
phenomenological combined gapped and power-law ansatz:
%
\begin{flalign}
T_1^{-1}(T) & =C_1(H)\exp\left(-\Delta/T\right) + C_2(H)(T-T_c)^\beta . & \label{mainfit}
\end{flalign}
%
%
Here, $C_1$ and $C_2$ are the weighting factors of the two contributions, $\Delta$ is the gap, and $T_c$ and $\beta$ are the critical temperature and
the exponent of the power-law contribution, respectively. Possible $T$-dependences of the prefactors $C_i$ remain unknown so far and have been not
considered here. Then the  \slrr$(T)$ dependences for all applied fields can be consistently modelled yielding a good description of the experimental
data as shown in Fig.~\ref{T1}. The field dependences of the parameters of Eq.~(\ref{mainfit}) are plotted in the Suppl., Fig.~\ref{T1fitparam}. The
critical power-law behavior of \slrr$(T)$ at 3\,T which sets in at $T \lesssim 7$\,K is fully consistent with the growth of the short-ranged
incommensurate correlations reported for this temperature regime in Ref.~\cite{Dutton12}. The fit requires a very small $T_c \sim 0.2$\,K suggesting
that the 3D long-range magnetic order, if present at all, is shifted to very low temperatures. At 9\,T, however, $T_c$ is pushed up to $\sim 1.5$\,K
indicating the proximity to a new, field-induced magnetic state that has been revealed in the specific heat data in Ref.~\cite{Dutton12}. Further
increase of the field up to 12\,T yields a reduction of $T_c$ down to $\sim 1$\,K, again consistent with the fading of the magnetic anomaly in the
specific heat \cite{Dutton12}. Interestingly, the best agreement with experiment for $H = 12$\,T requires a non-zero gap value $\Delta \sim 2$\,K in
the first term of Eq.~(\ref{mainfit}) implying the contrasting gapped and critical power-law contributions to \slrr$(T)$ with $\Delta > 0$ and $\beta
< 1$. By crossing the IBP $H_{\rm c1} \approx 13$\,T the critical growth of \slrr$(T)$ by lowering $T$ turns into a decay corresponding to the sign
change of the exponent $\beta$ (Fig.~\ref{fitgap}, inset). Concomitantly the weight $C_1$ of the gapped term in Eq.~(\ref{mainfit}) increases on
expense of the decreasing weight $C_2$ of the power-law term. At the same time $\Delta$ increases non-linearly (Fig.~\ref{fitgap}).
\\

\noindent\underline{
{\it Exclusion of an ordinary spin gap}}. In principle, there is a variety of conventional reasons for the opening of a gap in the spin excitation
spectrum of a quantum magnet exposed to an external field. Generally, above saturation where the spins are fully polarized, all excitations acquire a
gap that linearly scales with $H$. In the frustrated $J_1$(FM)\,-\,$J_2$(AFM) chain a two-magnon excitation is expected to have the lowest energy
(see, e.g., \cite{Zhitomirsky10}). In the case of \LCSO, neglecting the non-linearity of the $\Delta(H)$
dependence in the crossover region, the
increase of $\Delta$ amounts to $\Delta /H = 1.56$\,K/T (Fig.~\ref{fitgap}). This slope is in accord with the Zeeman energy of the flip of a single
spin, i.e.\ a one-magnon excitation, which with the $g$-factor $g = 2.18$ obtained in the ESR experiment would amount to $g\mu_{\rm B}/k_{\rm B} =
1.47$\,K/T. Correspondingly, the two-magnon slope should be $\sim 3$\,K/T. Anyhow, we note that it would be unreasonable to identify $H_{\rm c1}
\approx 13$\,T as an effective saturation field. Our measurements of the static magnetization at very low $T$ did not reveal a saturation of $M(H)$
even in 20\,T [Fig.~\ref{FigESR}(a)]. This finding is supported by our DMRG results (see below) showing that in a situation of the symmetric exchange
anisotropy present in \LCSO, there is no well defined saturation field at all in a literal sense, i.e.\ the full saturation at $T=0$ is achieved only
asymptotically [Fig.~\ref{mag_nematic}(a)].

Another possible reason for a field induced gap could be the presence of staggered antisymmetric Dzyaloshinskii-Moriya (DM) interactions. Due to the
low crystallographic symmetry, various DM interactions are generally allowed in \LCSO\ (see below). Their magnitude in \LCSO\ can be judged from the
ESR data, because ESR is very sensitive to magnetic anisotropies. Assuming that the strongest antisymmetric DM coupling is present for the
intra-chain NN bond with the DM vectors perpendicular to the Cu chain (see Fig.~\ref{structure} and Suppl.), a strongly anisotropic gap should open
for fields applied along the chain \cite{Oshikawa97}. Such a gap results in a shift of the ESR signal for this field direction proportional to $H^3$
at low temperatures $T<J/k_{\rm B}$ \cite{Oshikawa99,Oshikawa02}. However, experimentally the frequency vs.\ magnetic field dependence of the ESR
signal is linear within the error bars  over a broad field range [Fig.~\ref{FigESR}(c)] which suggests that the staggered DM component of the
antisymmetric exchange is small in \LCSO. The uniform component of the DM exchange can give rise to a field independent anisotropic gap which for
certain field orientations may yield a splitting of the ESR signal \cite{Povarov11,Smirnov15}. Such a fine structure of a powder ESR spectrum of
\LCSO\ is indeed found at high fields [Fig.~\ref{FigESR}(c)]. Its extend of the order $\approx \pm 30$\,GHz\,$=\pm 1.5$\,K could give then the energy
scale of the uniform DM component which is of a percent order of the isotropic and symmetric
anisotropic exchange couplings as estimated from the magnetization data (see below).\\

\noindent\underline{\it Evidence for spin-nematicity}. Ruling out the above discussed ordinary grounds for the field-dependent spin gap in \LCSO\
enables us now to focus on a possible, more sophisticated reason for the gap opening by approaching the IBP $H_{\rm c1} \approx 13$\,T from the
low-field side. According to the proposed theoretical precursor phase diagram of the isotropic frustrated $J_1-J_2$ spin chain
\cite{Vekua07,Hikihara08,Sudan09}, sufficiently high magnetic fields yet smaller than the saturation field induce a multicomponent spin liquid
including multi-magnon bound states. The two-magnon bound state ($p = 2$), corresponding to a precursor of a quadrupolar (spin-nematic) phase with a
finite four-spin correlation function $\langle S^{+}_{j}S^{+}_{j+1}S^{-}_{0}S^{-}_{1}\rangle$ is the simplest one. In the presence of detrimental for
the bound states interchain coupling  a collinear and incommensurate quasi long-range ordered SDW$_2$-phase is more and more stabilized at the lower
side of this field region since $\langle S^{z}_{j}S^{z}_{0}\rangle$ is the slowest decaying correlator. Above a certain crossover field the
quadrupolar $\langle ++--\rangle$ correlations might become nevertheless dominant yielding a competing quasi long-range ordered pronounced
spin-nematic state at the higher field side \cite{Vekua07,Hikihara08,Sudan09}. In both SDW$_2$ and the spin-nematic parts of the quadrupolar TL
liquid, $\langle S^{+}_{j}S^{-}_{0}\rangle$ is expected to be gapped as demonstrated qualitatively for the special quasi-2D isotropic model case at
$T=0$ \cite{Starykh14} which might probably hold in the case of relevant 3D interchain couplings at finite $T$, too. Finally, at very low fields
magnon bound states as well as the collinear SDW fluctuations/order will be suppressed. Instead, a vector chiral order, which typically arises in a
spin chain due to magnetic frustration somewhat modified quantitatively by possible DM couplings, turns out to be the ground state. In this phase the
gap closes and the transverse $\langle S^{+}_{j}S^{-}_{0}\rangle$ correlation becomes dominant.

Since quadrupolar correlations do not generate any fluctuating fields at a nuclear site,
Sato {\it et al.}~\cite{Sato09,Sato11} have proposed an
indirect way to identify the quadrupolar phase of the isotropic $J_1$(FM)\,-\,$J_2$(AFM) chain and to distinguish between its SDW$_2$ and the
spin-nematic dominated regions. It is predicted that \slrr\ due to longitudinal $\langle  zz \rangle$ correlations should follow the power law $\sim
T^{2\kappa-1}$, where $\kappa$ is the TL parameter. In the SDW$_2$ precursor phase $\kappa<1/2$ and \slrr\ diverges with decreasing temperature
whereas in the spin-nematic state $\kappa > 1/2$ and \slrr\ decays as $T \rightarrow 0$. In both regimes transverse $\langle +- \rangle$ correlations
yield a gapped contribution to \slrr$\sim \exp (-\Delta/T)$.

It is reasonable to attribute incommensurate spin correlations observed in \LCSO\ as well as a weak magnetic anomaly in the specific heat at low
fields \cite{Dutton12} with the onset of a conventional short-range ordered vector chiral  phase. The proximity to this phase is reflected in an
increase of the \li\ relaxation rate at low $T$ due to the growth of $\langle  +- \rangle$ correlations. From the fit with Eq.~(\ref{mainfit}) a
long-range vector chiral order due to interchain coupling could be realized only at very low temperatures $T_c \sim 0.2$\,K and in fact was not
observed down to 0.1\,K \cite{Dutton12}. A new field-induced magnetic phase at higher fields and higher temperatures yielding a strong anomaly in the
magnetic specific heat \cite{Dutton12} can thus be naturally ascribed to the short-range ordered  SDW$_2$ phase. In this regime the observed strong
enhancement of \slrr$(T)$ should be due to the dominant longitudinal $\langle  zz \rangle$ correlations which according to the modelling of the 9\,T
data with Eq.~(\ref{mainfit}) should yield a long-range SDW$_2$ order below $T_c \sim 1.5$\,K \cite{Kuehne17}.

Further increasing the field up to 12\,T yields a weakening of the
$\langle  zz \rangle$ power law contribution on which background a
gapped $\langle  +- \rangle$ contribution to the \slrr$(T)$ dependence
with $\Delta \sim 2$\,K becomes distinguishable. This
clearly suggests the destabilization of the SDW$_2$ state which is also
reflected in the decreased value of $T_c \sim 1$\,K in the model dependence
(Suppl., Fig.~\ref{T1fitparam}).

The vanishing of the power-law contribution at the critical IBP $H_{\rm c1} \approx 13$\,T signals then a crossover to the distinctive spin-nematic
state with dominant quadrupolar $\langle  ++-- \rangle$ correlations. In this regime the $\langle  zz \rangle$ correlations are decaying with
lowering $T$ which corresponds to the sign change of the power-law exponent $\kappa$ \cite{Sato09,Sato11}. This is indeed the case for \LCSO\
(Fig.~\ref{fitgap}, inset). The now decaying power law contribution to \slrr\ loses progressively its weight with increasing field whereas the gapped
contribution becomes dominant (Suppl., Fig.~\ref{T1fitparam}). Indeed, as it has been emphasized in Ref.~\cite{Starykh14} the gapped excitation
spectrum is a distinct feature of the spin-nematic state of the weakly coupled 1D-chains with only a weak soft mode in the longitudinal $\langle zz
\rangle$ channel \cite{Syromyatnikov12,Starykh14}. Thus, regardless of the specific details the above analysis points at a rather broad stability
range of the spin-nematic state in \LCSO\ above the narrow crossover region at the IBP field $H_{\rm c1} \approx 13$\,T.

Owing to small and/or frustrated interchain couplings in \LCSO, long-range order of any kind is likely to be suppressed by quantum fluctuations to
very low $T$. Therefore, the distinct state around $H_{\rm c1}$ identified in the NMR experiments in the accessible temperature range should be
considered as the TL quadrupolar {\it liquid} with dominant SDW$_2$ correlations for $H \lesssim H_{\rm c1}$ and with dominant spin-nematic
correlations for $H\gtrsim H_{\rm c1}$. In a recent 2D-model the SDW$_2$ state is shown to dominate over a broad magnetic field range squeezing a
nematicity region very near to the saturation field \cite{Starykh14}. In contrast, the SDW$_2$ liquid phase in \LCSO\ appears to be confined to a
restricted field range below $H_{\rm c1}(H)$ at lower $T$ not covered by the present NMR study, giving much place to the competing pronounced spin
nematic state in an extended field range above $H_{\rm c1}$. This may be related to the specific interchain couplings in \LCSO\ and the detrimental
influence of DM couplings favoring noncollinear order and thus destabilizing any collinear SDW$_p$, ($p \geq 2 $) state. In fact, a weak interchain
coupling is favorable for a pronounced multipolar state \cite{Nishimoto15} but detrimental for any dipolar SDW-state or spiral state since the
Ne{\'e}l temperature scales usually with coherent interchain coupling which suppress the detrimental thermal fluctuations. Thus possible structural
disorder, the presence of DM interactions and strongly frustrating interchain couplings allowed by the low-symmetry (Fig.~\ref{structure}) are also
not favorable for any magnetic dipolar ordering. The mentioned NN symmetric exchange anisotropy stabilizes the nematicity as will be shown below (see
also Ref.~\cite{Nishimoto15}). All these circumstances are obviously favorable for a stabilization of a nematic state in a broad region above $H_{\rm
c1}$ as depicted in the schematic phase diagram of \LCSO\ in Fig.~\ref{pd}. Certainly, there must be also a second ''upper '' critical field $H_{\rm
c2}$ framing the stability region of the strong nematic state in \LCSO. This calls for further experimental studies of \LCSO\ at higher fields and
also at lower temperatures beyond the scope of the present work.

\noindent{\bf Theory.} \underline{\it Band structure calculations}. As in other related materials (see., e.g.,
Refs.~\cite{Drechsler07,Buettgen14,Prozorova15,Willenberg16}), the edge-sharing geometry of the CuO$_4$ plaquettes in the CuO$_2$ chains in \LCSO\
(Fig.~\ref{structure}) is expected to give rise to the usual frustrated magnetism due to the presence of oxygen mediated frustrated AFM NNN
intra-chain couplings. The nearly 90$^o$ Cu-O-Cu bond angle, i.e.\ 93$^o$,  points to FM NN intra-chain interactions due to the presence of a sizable
direct FM coupling $K_{pd}\sim 100$\,meV between two holes residing on neighboring sites with Cu 3$d$ and O $2p$ orbitals and a significant
compensation of the AFM NN superexchange contributions.

We have performed DFT and DFT+$U$ bandstructure calculations with the aim to understand (i) the amount of interchain couplings and (ii) the magnitude
of the intra-chain couplings. With respect to (i) we have analyzed the dispersion of bands and found pronounced 1D van Hove singularities near the
Fermi level. Thus, we have confirmed the nearly 1D behavior of \LCSO. Then, in general, the exchange coupling strength can be estimated simply by the
AFM contribution $J_1=4t_1^2/U_{\rm eff}$, where $U_{\rm eff} \sim \Delta_{pd} \approx 3$ to 4~eV is the effective Coulomb repulsion within a {\it
single}-band type approach for the Cu-sites with NN transfer integral $t_1$. The frustrating $J_2$ is measured by the analogous expression using the
NNN transfer integral $t_2$ instead ignoring a much weaker direct FM contribution as compared to that of the NN bond ($K_{pd} \gg K_{pp}$) and the
small hole occupation at O 2$p$ orbitals. Note that applying such a simple model to charge transfer insulators as cuprates, one is left with an
effective onsite repulsion $U$ of the order of the Cu 3$d$-O 2$p$ onsite energy difference which is significantly smaller than the $U_d \sim $~5.5~eV
at Cu sites employed in the DFT+$U$ calculations or in more sophisticated five-band $pd$ Hubbard models \cite{Malek08,Monney16} to be considered for
to the case of \LCSO\ elsewhere. To check this simple first approach, we have determined the Cu-Wannier functions which contain also the essential O
2$p$ contributions. Their tails point to the important coupling directions. In fact, a closer inspection of the crystal structure reveals
nonequivalent "left" and "right" NN intra-chain bonds (Fig.~\ref{structure}). This gives rise to alternating NN transfer integrals ($t_1 \neq t'_1$).
The one-band fit results in the following transfer integrals (given in meV): $t_1=-95.51,\quad t'_1= 56.44, \quad t_2=56.96, $ and $t_3= -11.88,
\quad t'_3=-16.57$\,. Then the mean NN transfer integral $\bar{t}_1$ would provide an AFM superexchange contribution to $J^e_1$ for an equidistant
chain of about 67~K which for a typical $J^e_1$ of about $-80$\,K like in linarite (see, Refs.~\cite{Willenberg16,Rule16} and references therein)
corresponds to an FM contribution of $-147$\,K. As a result we arrive at a sizable splitting of the two NN exchange integrals: $J_1 \approx -160$~K
and $J_1' \approx -90$~K, whereas $J_2 \approx $~37.6~K, only. Thus, within a 1D picture we are left with a dominant FM total NN coupling and an
unrenormalized mean frustration parameter $\bar{\alpha} = J_2/[(J_1+J_1')/2] \sim 0.3$. This value is close to that ($\alpha=0.28$) estimated from
the fitting of the observed magnetization curve by the DMRG calculations
(see below). \\

\noindent\underline{\it Symmetry analysis and the role of DM interactions}.
The crystal structure of \LCSO\ is described by the
polar space-group Cmc2$_1$ \cite{Dutton12}. The low symmetry implies modifications of the standard $J_1$-$J_2$ spin-model to
describe the chains of the edge-shared CuO$_4$ square-like plaquettes runnning in $a$-direction. More details of our symmetry
analysis are given in the Supplement. In particular, because of the low symmetry, antisymmetric  Dzyaloshinskii-Moriya (DM)
interactions are allowed for the NN bonds along the spin-chains, \(E_{D} = \mathbf{D}_{\nu}\cdot (\mathbf{S}_i \times
\mathbf{S}_{i+\nu})\) and have {\it both} a homogeneous and a staggered component. These microscopic antisymmetric exchange
interactions are caused by the relativistic spin-orbit interactions and compete with the isotropic exchange interactions.

To fit the experimental data for \LCSO, Dutton {\it et al.} \cite{Dutton12} assumed in their model the presence of this exchange
anisotropy for the NN bond, while neglected the anisotropic terms {\it linear} in $|\mathbf{D}_{\nu}|$. However, this could yield
an unrealistic picture for the basic magnetic couplings in  \LCSO. The determined strength of the effective anisotropic coupling
constant is large and would imply an unusual order of magnitude $|\mathbf{D}_1|/|J_1| \sim 1$. The presence of the relativistic
antisymmetric exchange in crystals belonging to the crystal classes 2mm or $C_{2v}$ may give rise to two rather different states
\cite{Bogdanov02}: weak ferromagnetism or more generally canting of spins with a net magnetization can be derived by the
staggered DM couplings. Besides, the acentric crystal structure also allows for the presence of `{\textit{inhomogeneous DM
couplings}',\cite{Bogdanov02} that derive from the homogeneous part of the DM interaction. It is known that this type of
couplings can suppress any
long-range ordered states and may be related to the absence of
magnetic ordering down to $\sim 0.1$~K in \LCSO.\\

\noindent\underline{\it DMRG-calculations}. The main aim of this part
is to present an analysis  of a novel anisotropy mechanism based on the
low-symmetric NN exchange anisotropy, which in addition to the $J_1$-$J_2$
frustration, stabilizes a nematic phase in a moderate high-field region to
be specified below. We present a first brief analysis also of the effect
of weak homogeneous and staggered NN DM couplings which were found not to
destroy the nematicity although some weakening has been observed.

\begin{figure}[t]
\centering
\includegraphics[clip,scale=0.6]{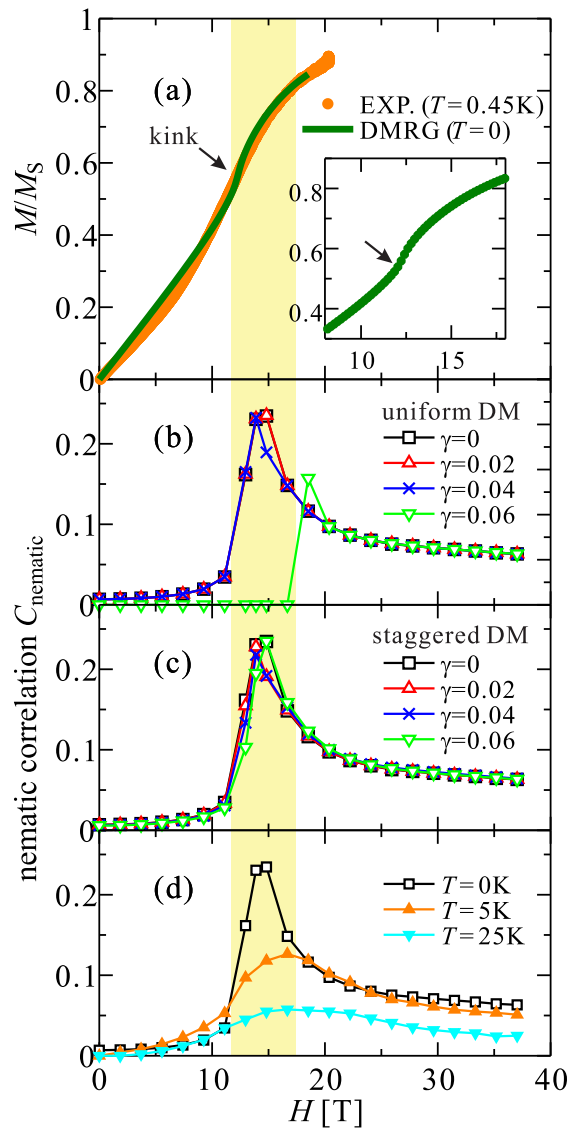}
\caption{(a) Experimental magnetization curve
at $0.45$\,K and theoretical magnetization curve calculated for $(J_1^x,J_1^y,J_1^z)=(-1.07,-0.99,-1)$ and $J_2=0.28$. Inset: enlarged figure around
kink in the theoretical curve. (b) Nematic correlation for a homogeneous DM coupling $\gamma=0$, $0.02$, $0.04$, and $0.06$. (c) the same for a
staggered DM coupling. (d) Temperature dependence of the nematic correlation. The shaded area depicts the field range where a spin gap was observed
in the NMR experiment.}\label{mag_nematic}
\end{figure}
Fig.~\ref{mag_nematic}(a) shows the magnetization curve $M(H)$ measured at $T=0.45$\,K. Noteworthy, a full saturation is not reached even in the
highest accessible field of 20\,T. Although this is reminiscent of a typical feature of $M(H)$ at high temperature, the observation temperature is
now low enough to prevent a significant finite-temperature effect. To provide a reasonable explanation for this feature, we introduce a 1D frustrated
Heisenberg model with an $xyz$ exchange anisotropy and a magnetic field $H$ along the $z$ axis. The Hamiltonian is then given by
\begin{eqnarray}
\nonumber \mathcal{H} = \sum_{i,\gamma=x,y,z} J_1^\gamma S^\gamma_iS^\gamma_{i+1}
&+& J_2 \sum_i \mathbf{S}_i \cdot \mathbf{S}_{i+2} \\
&+& H \sum_i S_i^z,
\label{hamxyz}
\end{eqnarray}
where $J_1^\gamma$ and $J_2$ are the NN FM and the NNN AFM exchange
couplings, respectively,
and $S^\gamma_i$ is the $\gamma$-component of the
spin-operator
$\mathbf{S}_i$. When $H \gg J_1^\gamma, J_2$, by taking the fully
polarized FM state as non-perturbative state,
its
energy is lowered by
$\Delta E = (J_1^x-J_1^y)^2/[32(H-J_1^z-J_2)]$ through the second-order
process of an individual double spin-flip. Therefore, the magnetization
behaves like $M_S-M \propto (J_1^x-J_1^y)^2/H$ at high fields and saturates
its maximum value $M_{\rm S}$ only at $H=\infty$, only. We have calculated the
magnetization curve using the DMRG technique. By fitting the experimental
curve, we have found a possible parameter set: $J_1^z=-546$\,K,
$(J_1^x/J_1^z,J_1^y/J_1^z)=(1.07,0.99)$, and $J_2=153$\,K. Note that these
numbers are effective values in the 1D limit, which can be significantly
different from the bare values of NN and NNN exchange couplings determined
by the DFT+$U$ calculations. Other contributions such as the interchain
and longer-range exchange couplings are renormalized into them. As a related
similar example we refer the reader to linarite \cite{Willenberg16,Rule16}.

More interestingly, an exotic nematic state is established by the
$xyz$ exchange anisotropy \cite{remarkphase}. The $xy$ components of the first term
of Eq.\ (\ref{hamxyz}) can be divided into an exchange
term $\frac{J_1^x+J_1^y}{4}(S^+_iS^-_{i+1}+h.c.)$ and a double spin-flip term
$\frac{J_1^x-J_1^y}{4}(S^+_iS^+_{i+1}+h.c.)$. The latter seems to create an
attractive interaction among the parallel spins. Therefore, a 2-magnon
bound state, i.e., a nematic state, may be naively expected at high field \slm{s}
in the presence of $xyz$ exchange anisotropy. To check this possibility, we
have calculated the nematic correlation function as an indicator of magnon pairing
\begin{eqnarray}
\mathcal{C}_{\rm nematic} &=& \langle S^-_iS^-_{i+1} \rangle - \langle S^-_iS^-_{i+\infty}
\rangle \nonumber \\
&\equiv &
\langle S^+_iS^+_{i+1} \rangle - \langle S^+_iS^+_{i+\infty} \rangle
 .  \\
 \nonumber
\label{nop}
\end{eqnarray}
Note that this correlation vanishes for lacking $xyz$ exchange anisotropy,
 because there is no overlap between different $S^z$ sectors. The nematic
correlation for our parameter set is plotted as a function of
 $H$ in Fig.~\ref{mag_nematic}(b). It is significantly enhanced just above the
kink position of the theoretical magnetization curve. This field range with the enhanced correlation agrees well with that region where the spin gap
has been experimentally observed, namely, in between $H=13-16$\,T. A similar nematicity scenario has been proposed in our recent work devoted to
linarite \cite{Willenberg16}, but there yet not fully confirmed experimentally due to a phase separation and other experimental and physical
difficulties. Furthermore, we have also studied the effect of additional uniform or staggered DM couplings allowed by the crystallographic symmetry
as mentioned above $\mathcal{H}_{\rm DM}=\sum_i \mathbf{D} \cdot (\mathbf{S}_i \times \mathbf{S}_{i+1})$ with $\mathbf{D}=(0,0,\gamma)$. As a result
we found that the nematic state is {\it hardly} affected by a weak DM coupling for $\gamma<0.05$.
The effect of a staggered DM
interaction is even weaker than that of a uniform one. For simplicity, the dimerization of the NN exchange couplings, suggested by the DFT, was not
taken into account in the present DMRG calculations. However, the stability of the nematic state is mostly related to the magnitude of the exchange
anisotropy and is less affected by the dimerization. Also, the $T$-dependence of the correlation is plotted in Fig.~\ref{mag_nematic}(d). We can see
that the sharp enhancement of the correlation around $H=13$\,T at $T=0$ disappears for higher $T$ which points to existence of the mentioned above
upper critical field.

Thus, we are confronted with a somewhat unusual situation: the pronounced spin gap and the strongly enhanced nematic correlations are, in a literal
sense, not the result of the occurrence of a novel order parameter associated with symmetry breaking due to a second order phase transition from a
high temperature and low-field para-phase, since at low fields the nematic order at $T=0$  already exists albeit at a low level. Instead, based on
our calculations and in accord with the experimental data, we suggest a crossover transition from a weak nematic state in a narrow field range at
about 13\,T to a pronounced nematic state up to at least 16\,T to 20\,T to be followed by a broad field range where it decreases again
(Figs.~\ref{mag_nematic} and \ref{pd}).

Such transitions without a symmetry change of the macroscopic order parameter  are reminiscent of mesoscopic liquid-liquid transitions in ordinary
liquids such as, for instance, in phosphorus and water governed by the {\it quantitative} change of a correlation function, only
\cite{Katayama00,Kurita04,Tanaka13}. With increasing $T$ these changes are smeared out and  the consequences of the suppressed nematic order
parameter are difficult to be observed. In such a complex situation further experimental and theoretical studies beyond the scope of the present work
 are necessary to refine the parameters of our proposed model and to take into account explicitly the weaker couplings and modifications
suggested by the real structure including also impurities or defects.
\\

\begin{figure}[t]
\centering
\includegraphics[width=0.8\columnwidth]{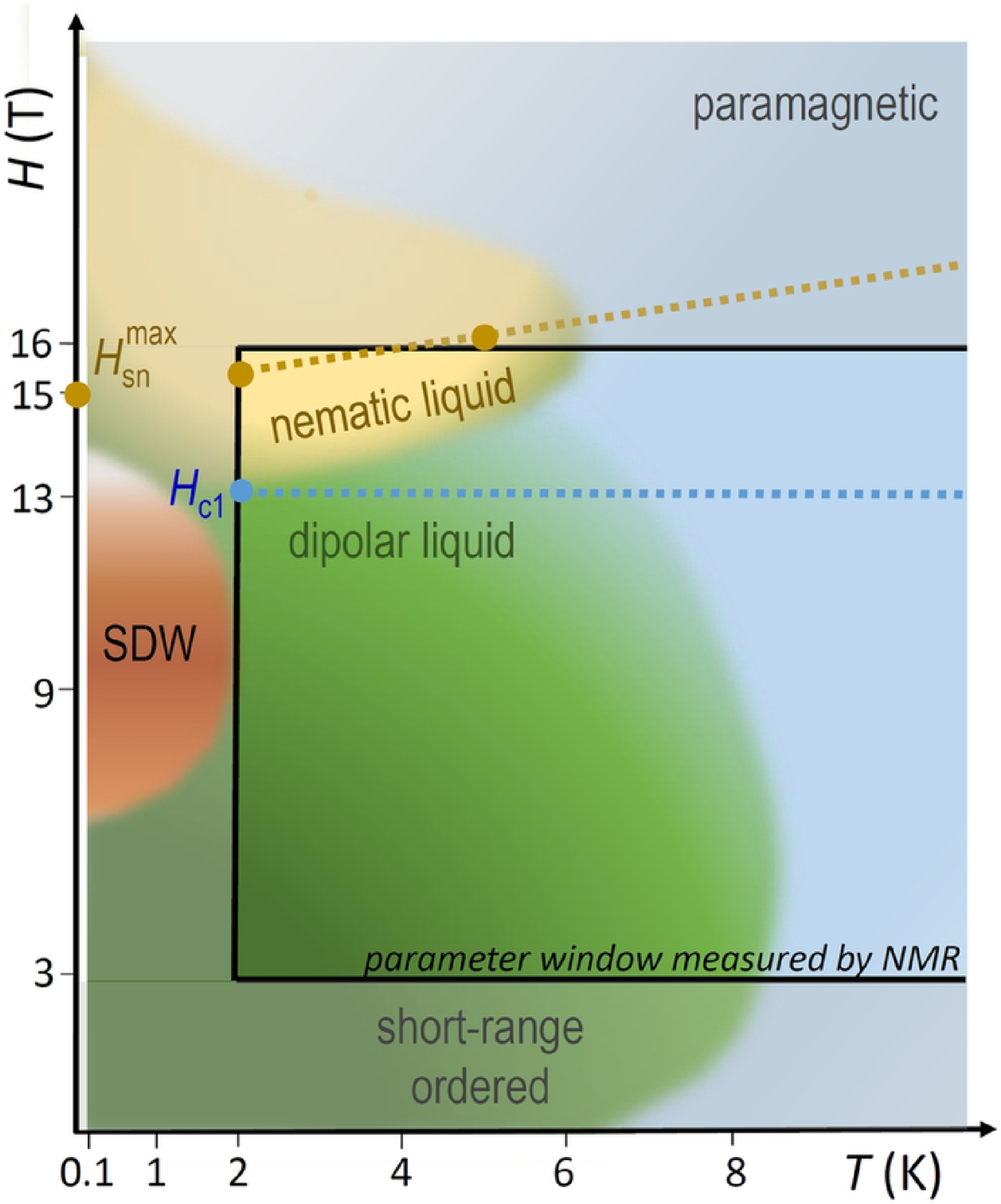}
\caption{Schematic phase diagram of \LCSO . Blue, dark green\slm{,} and dark red regions reproduce approximately  the diagram of Dutton {\it et al.}
based on the analysis of the specific heat and magnetization data (Fig.~3 in Ref.~\cite{Dutton12}). The dark red area is suggested to present an
anomalous SDW phase, whereas the dark yellow area depicts an envisaged stability region of the proposed nematic state. The region measured by the NMR
in the present work is marked by the black rectangle. The blue dashed line denotes the isosbestic field $H_{\rm c1}$ (cf. Fig.~\ref{Figisos}). The
brown closed circles labelled $H_{\rm sn}^{max}$ connected with the dashed line depict the field of the maximum of the spin\slm{-}nematic correlation
function as found in the DMRG analysis (cf.\ Fig.~\ref{mag_nematic}). } \label{pd}
\end{figure}

\noindent{\bf Summary}\\
In conclusion, we have presented experimental and theoretical evidence for the occurrence of a distinctive spin-nematic state in the frustrated
anisotropic spin chain cuprate \LCSO\ . This state emerges above an isosbestic point $H_{\rm c1} \approx 13$\,T detected in the field dependence of
the $^7$Li NMR relaxation rate \slrr\ at low temperatures. The analysis of the temperature dependences of \slrr\ reveals that $H_{\rm c1}$ separates
a lower-field regime with strong enhancement of \slrr\ at low $T$ from a higher-field regime with a sharp decrease of \slrr$(T)$. The former is
ascribed to diverging longitudinal spin correlations typical for a multipolar SDW liquid whereas the latter is due to the power-law like decaying
longitudinal and gapped transverse spin correlations characteristic of the spin-nematic liquid. Theoretical analysis justifies the occurrence of this
"hidden" spin-nematic state in \LCSO\ in an extended field range above $H_{\rm c1}$. A broad range of stability of this spin-nematic state unexpected
in the corresponding isotropic spin-chain model can be ascribed to the presence of exchange anisotropies. Indeed, as is found in the DMRG
calculations it can be due to a special low-symmetry symmetric exchange anisotropy which is reflected in the high-field magnetization data. The
missing SDW-type magnetic ordering at lower fields might be ascribed to some structural disorder and/or frustrated interchain interactions caused by
DM couplings allowed in this low-symmetry crystal structure in general. The small but finite DM coupling favoring a noncollinear spin arrangement
could be responsible for the suppression of the otherwise strongly competing anomalous collinear SDW$_2$ phase. On the other hand, according to
complete diagonalization and DMRG studies in high magnetic fields the presence of a weak DM interaction, uniform or staggered, identified in the
symmetry analysis of \LCSO\ and assessed with ESR, are not detrimental for nematicity. Merely an alteration of the two NN exchange couplings may
somewhat reduce the binding energy of two-magnon bound states as compared to equal NN bonds. The remarkable interplay of symmetric and antisymmetric
exchange anisotropies with sizable frustration is of general interest
in modern quantum magnetism and calls for deeper theoretical and experimental studies.\\

\noindent{\bf Methods}\\

 {\bf Sample synthesis and characterization.} Green polycrystalline sample
 of \LCSO\ was prepared through solid-state reaction using
stoichiometric mixture of dried Li$_2$CO$_3$ (99.98\,\%, Chempur), CuO (99.95\,\%, Aldrich) and Sb$_2$O$_5$ (99.9\,\%, Alfa Aesar). The mixture of
the precursor compounds was homogenized by grinding in a mortar and pestle,
followed by a 16\,h sintering at 700$^\circ$\,C. The sample was
subsequently grounded, pressed into pellets and fired at 1000$^\circ$\,C for
48\,h. Pure \LCSO\ phase was obtained after annealing of the pellets at
1050$^\circ$\,C for 24\,h under dried oxygen flow. Phase purity of the
products was assessed by powder X-ray diffraction, by using a STOE Stadi P
powder diffractometer with Mo K$_{\alpha 1}$ radiation. The diffractometer
is equipped with a curved Ge (111) monochromator and a 6$^\circ$ linear
position sensitive detector (DECTRIS MYTHEN 1 K detector). Powder x-ray diffraction data were analyzed with the Rietveld method using the FULLPROF in
the WinPlotR program package program \cite{Carvajal93}. The background was fitted using linear interpolation between selected points. The
March-Dollase model for preferred orientation was used in all of the refinements, and a pseudo-Voigt function was used as the peak-shape model. As
refinable parameters background, scale factor, half width, Caglioti variables (U, V, W), lattice parameters, asymmetries and the overall temperature
factor were allowed. Based on Rietveld analysis of the powder x-ray diffraction data (Fig.~\ref{FigXray}), the main phase is \LCSO\ 96.93(6)\,wt\,\%
(orthorhombic, Cmc21, $ a= 5.7493(1)\AA$, $b = 10.8828(2) \AA$, $c = 9.7429(1) \AA$). LiSbO$_3$ 2.36(5)\,wt\,\% (orthorhombic, Pnma, $a = 5.1756(4)
\AA$, $b = 4.9092(3)\AA$, $c = 8.4887(6) \AA$) and 0.72(1)\,wt\,\% CuO are two minor impurity phases. The amount of foreign phase is very similar to
the previous report in Ref.~\cite{Dutton12}.

\begin{figure}
\includegraphics[width=0.8\columnwidth]{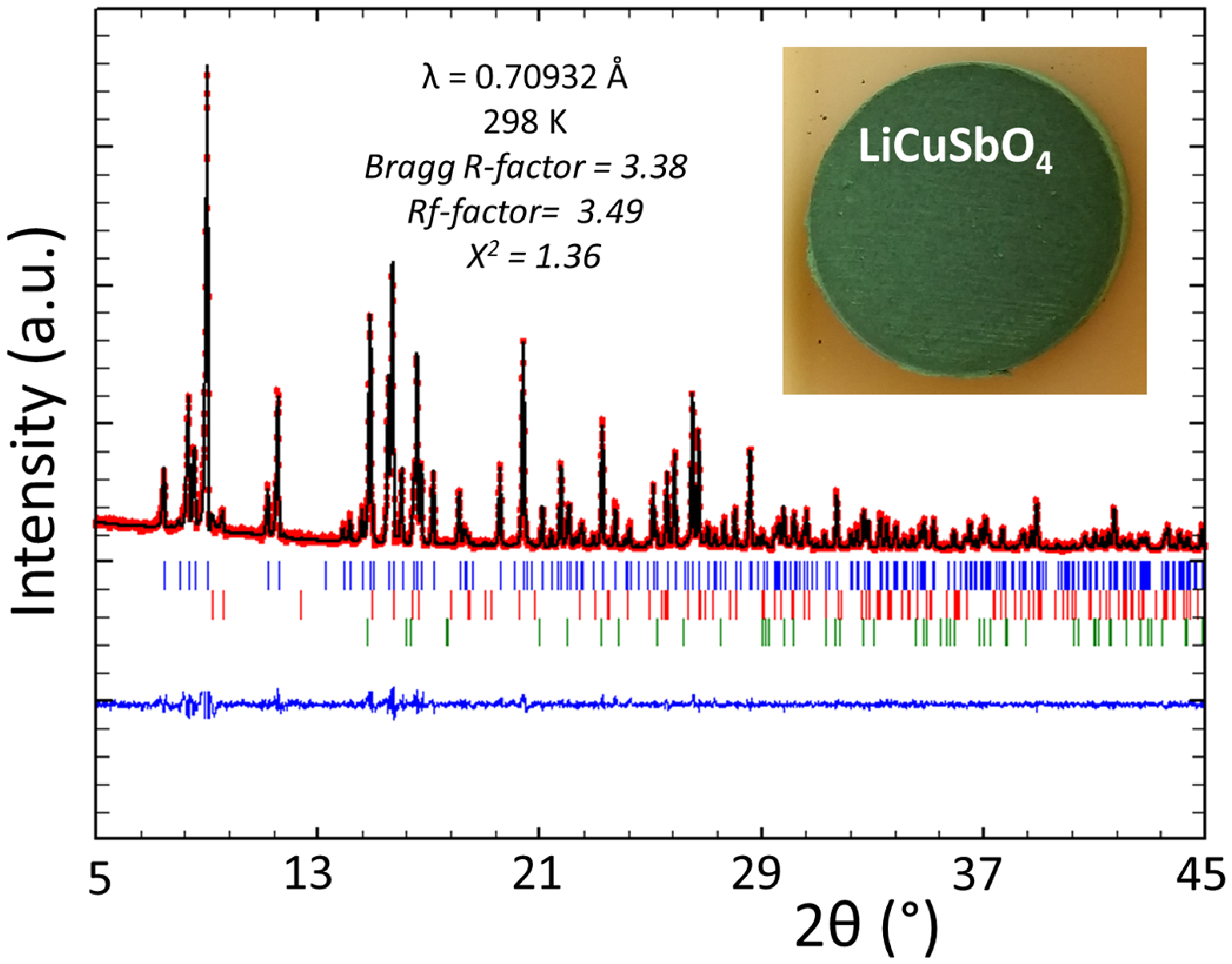}
\caption{Calculated and observed X-ray diffraction pattern of the Rietveld refinement for LiCuSbO4. The difference curve is shown
in blue; reflection positions are indicated by the vertical lines for LiCuSbO4 (blue) , LiSbO$_3$ [2.36(5)\,wt\,\%] (red) and CuO
[0.72(1)\,wt\,\%] (green) impurity phases. ($\lambda = 0.70932 \AA$, Bragg R-factor: 3.38\,\%; Rf-factor = 3.49\,\%; Bragg
R-factor = $\sum |I_{\rm ko} - I_{\rm kc}|/\sum I_{\rm ko}$; Rf-factor = $[(N - P)/\sum w_{\rm i}�y_{\rm io}2]^{1/2}$). The inset
shows optical image of a typical green pellet of polycrystalline sample of \LCSO\ after annealing in dried oxygen flow.}
\label{FigXray}
\end{figure}

{\bf Nuclear magnetic resonance.} NMR spectra were obtained with a Tecmag Apollo spectrometer and a standard sample probe from NMR Service GmbH. The
magnetic field has been applied by a 16 T Oxford Instruments superconducting magnet. Temperatures were regulated by a $^4$He variable temperature
insert (VTI). Temperatures below 4.2\,K were achieved by pumping on the VTI.
At high temperatures and small fields, Fourier transformations (FFT) of
the spin echo covered the whole spectral width. At lower temperatures, we swept the frequency and summed up the FFT's to obtain the complete
spectrum. The spectra at 15 T below 10 K have been obtained by field sweeps, and converting into frequency sweeps. This is easily possible due to the
negligible quadrupole interaction. We have confirmed the correctness of this procedure at higher temperatures. The spin lattice relaxation rate,
\slrr , has been measured by standard saturation recovery at the peak of the spectra. The nuclear magnetization, $M_0$, has been saturated by a train
of radio frequency pulses, before measuring the recovered nuclear magnetization, $M(\tau)$, depending on the time $\tau$ between the saturation train
and the spin echo sequence.

{\bf Electron spin resonance.} ESR spectra were measured with the Terahertz ESR Apparatus (TESRA-IMR) installed in the magnetism
division of Institute of Materials Research, Tohoku University \cite{Nojiri06}. As sources of the microwave radiation up to
450\,GHz conventional Gunn oscillators were employed. The signals were detected with an InSb detector. Pulse magnetic fields up
to 20\,T were generated with a solenoid magnet and a 90\,kJ capacitor bank. The sample temperature was regulated with a $^3$He
cryostat. Additional ESR measurements were performed at the IFW Dresden with a home-made multi-frequency high-field ESR
spectrometer at magnetic fields up to 16\,T and at frequencies $\nu$ up to 400\,GHz \cite{Golze06}. For the generation and
detection of the microwave radiation millimeter wave backward oscillators and an InSb bolometer from QMC Insruments Ltd., as well
as a millimeter wave network analyzer from AB Millimetre, have been used. DC magnetic fields were obtained with a solenoid
superconducting magnet from Oxford Instruments equipped with a $^4$He variable temperature insert.

{\bf Static magnetization.} Temperature dependence of the static magnetic susceptibility in fields up to 5\,T in the temperature range $T = 2 -
300$\,K was measured with the SQUID magnetometer from Quantum Design. Static magnetization in fields up to 20\,T was measured with a standard
inductive method using compensated pickup coils and a nondestructive pulse magnet (for details see Ref.~\cite{Nojiri04}). The sample was immersed
into liquid $^3$He to reach a temperature as low as 0.45\,K.\\

{\bf Density functional calculations.} Relativistic density functional (DFT) electronic structure calculations were performed using the
full-potential local orbital FPLO code \cite{fplo1,fplo2,fplo3}, version fplo14.00-49. For the exchange-correlation potential, within the local
density (LDA) and the the general gradient approximation (GGA) the parametrizations of Perdew-Wang \cite{PW} and Perdew-Burke-Ernzerhof \cite{PBE}
were chosen, respectively. Both exchange-correlation potential yielded essentially the same band structure. To obtain precise band structure
information, the final calculations were carried out on a well converged mesh of 4800 $k$-points (20x20x12 mesh, 1332  points in the irreducible
wedge of the Brillouin zone). For our calculations, we used the experimental crystal structure of Ref.~\cite{Dutton12}. However, to model the Li
split positions Li1a (4a), Li1b (4a) and Li2 (8b), we used averaged coordinates: Li1 (4a) 0.0 0.3610 0.6974 or the coordinates of the nearby high
symmetry position Li2 (4a) 0. 0.0002 0.2660; Li split positions have been successfully modeled this way for the related edge-sharing chain compound
LiZrCuO$_4$ \cite{LiZrCuO}.\\

\noindent{\bf Acknowledgments}\\

The authors thank C.\ Blum and L.\ Giebeler for assistance with the XRD measurements, and A. Wolter-Giraud, F. Hammerath, Y. Utz and A. Smirnov for
useful discussions. This work has been supported in part by the Deutsche Forschungsgemeinschaft through grants SFB 1143 (S.N.), KA 1694/8-1 (V.K.),
GR 3330/4-1 (H.-J.G.), through the Emmy Noether Programme in the projects WU595/3-1, and WU595/3-2 (S.W.). The work of E.V. and M.I. has been
supported in part by project RFBR 14-02-01194. V.K. gratefully acknowledges financial support during his stay in Sendai from the International
Collaboration Centre, Institute of Materials Research, Tohoku University, Sendai. S.-L.D and S.N.\ thank R.O.\ Kuzian, J.\ van den Brink, C.\
Agrapidis, A.\ Tsirlin, A.\ Zvyagin, P.\ Mendels, H.\ K\"uhne,
S.\ Zvyagin, M.\ Zhitomirsky, and O.\ Starykh for discussions and stimulating interest.\\

\noindent{\bf Author contributions}\\

H.-J.G., M.I and E.V. conducted the NMR studies. H.N., V.K., and A.A. conducted the ESR studies. H.N. performed the high-field magnetization
measurements. M.I.S. and S.W. synthesized and characterized the samples. S.N. performed the DMRG-calculations and invented the nematicity scenario
based on the special anisotropic  NN exchange. J.R. carried out the complete diagonalizations to study effects of the DM interactions. H.R. performed
the LDA-FPLO electronic structure calculations for \LCSO\ , invented its ($J_1, J'_1)-J_2$ model, and estimated the leading exchange couplings. U.R.,
H.R. and S.-L.D. made the symmetry and phenomenological analysis including the isosbestic point. L.S. calculated the hyperfine tensor. V.K., E.V.,
S.-L.D. and B.B. designed the project.  All coauthors participated actively in the discussion and selection of the various experimental and
theoretical results presented in this paper and worked out the final concept of the paper. V.K. and S.-L.D. wrote the paper with important
contributions
to all its parts from all other coauthors.\\

\noindent{\bf Additional information}\\
\noindent{\bf Competing financial interests:} The authors declare no competing financial interests.

\nocite{apsrev41Control}
\bibliography{LiCuSbO4_references}

\begin{thebibliography}{74}
\bibitem [{com({\natexlab{a}})}]{commentMourigal12}%
  \BibitemOpen
  \href@noop {} {} \bibinfo {note} {{F}or completeness we
  note, that a suggested "bond-nematic" phase with a coexisting collinear
  SDW$_2$-magnetic structure and nematic fluctuations proposed in
  Ref.~\cite{Mourigal12} has been arguably questioned in Ref.~\cite{Starykh14}
  based on the analysis of an isotropic quasi-2D field-theory model for
  interacting chains. The very broad phase below $H_{\rm sat} \sim$ 43~T was
  ascribed instead to the mentioned above SDW$_2$-phase, only.}\BibitemShut
  {Stop}%
\end{thebibliography}

\clearpage

\noindent{\bf Supplement}\\

\noindent{\bf Details of the exchange and kinematic interactions as derived from the band structure calculations}  The two tiny third neighbor
interactions $J_3 \sim $ 1.6 to 3.2~K, respectively, were both ignored in the DMRG calculations for the sake of simplicity, i.e.\ first of all to
restrict the number of model parameters. Note that the NN exchange integrals may have a margin of error about a few ten K due to several
uncertainties like in $U_{\rm eff}$ and a more accurate estimation is left for future work. However, the alternation of the NN exchange integrals is
most likely, a rare situation so far only met for the celebrated spin-Peierls compound CuGeO$_3$ \cite{Hase93} with two AFM NN couplings and probably
with two ferromagnetic ones for Rb$_2$Cu$_2$Mo$_3$O$_{12}$ \cite{Hase04,Agripidis16}, only. The interchain transfer integrals were derived employing
the calculated Wannier functions to reproduce the full DFT band dispersion of the band complex near the Fermi energy. Their overlapping tails are
smaller by an order of magnitude $\sim$ 10~meV as compared to the NN inchain values giving this way rise to still smaller frustrated AFM
contributions of the order of 1\,K,only, (Fig.~\ref{structure}). Since there are no reasons to expect significant FM contributions the resulting
total interchain couplings are expected to be extremely small, too. This qualitative estimate might explain the strongly supppressed SDW and spiral
phases and most importantly will only weakly suppress the nematic state \cite{Nishimoto15}.\\

\noindent{\bf Detailed symmetry analysis: Non-centrosymmet-ric crystal structure from the class $C_{2v}$: generalities } The crystal structure of
LiCuSbO$_4$ has been described in the acentric (polar) space-group Cmc2$_1$ (No~36, point symmetry $C_{2v}$) \cite{Dutton12}. The low symmetry allows
that DM interactions,
%
\begin{equation}
E_{D} = \mathbf{D}_{ij}\cdot (\mathbf{S}_i \times \mathbf{S}_j) ,
\end{equation}
%
can occur for {\textit{all}} bonds between two Cu-sites. A long-range-ordered magnetic state (or the hypothetical mean-field magnetic order) in these
materials may appear as a basically AFM spin-pattern that is twisted into long-period textures. A 1D texture is known as a "Dzyaloshinskii spiral". A
monodomain-state of such a spiral has a spatially fixed rotation axis w.r.t.\ to its propagation direction and a fixed (chiral) sense of rotation. In
the crystal classes $C_{nv}, n=2,3,4, 6$ the possible propagation directions $\mathbf{p}$ are perpendicular to the unique axis of the crystals
$\mathbf{c}$. The propagation direction in the ground-state is determined by weaker anisotropies of the magnetic system. The rotation axis for the
spins is transverse to this propagation axis and also is in the plane perpendicular to the crystallographic axis. Thus, the staggered vector rotates
in a cycloidal manner within the plane spanned by the propagation direction $\mathbf{p}$ and the axis $c$. This slow rotation of the primary
order-parameter components is accompanied by a (weak-FM) spin-density wave, similar to magnets from crystal classes ($C_{nv}, n=3,4, 6$) analysed in
Ref.~\onlinecite{Bogdanov02}.

In general, Lifshitz invariants in several spatial directions do {\it frustrate} long-range ordering (LRO) as these terms act like a frozen gauge
potential background on the incipient ordering and enhances the impact of fluctuations to destroy LRO as has been described for a wide range of
different systems from cholesteric liquid crystals \cite{Meiboom81} to chiral magnets \cite{Wilhelm12}. In a paramagnetic but spin-liquid state of an
AFM system with $C_{2v}$ symmetry,  a possible kind of correlated low-energy excitations would consist of skyrmionic modes. These are localized
excitations (possibly in the shape of ellipsoidal staggered spin-states) that are twisted in the two basal plane directions. Other possible
excitation modes would be 1D kink-like solitonic units with an envelope function of the order-parameter that restricts their spatial extension on a
paramagnetic background.

In both kinds of such non-linear excitations, the magnitude and direction of the AFM correlations within such a fluctuating excitation are
intertwined and cannot be separated \cite{Toledano87,Yamashita87,Mukamel85,Wilhelm12}. For a fully established 3D spin-order, the Lifshitz-type
invariants for the vector components of a magnetic ordering mode have spatial gradients in the base-plane (perpendicular to the $n$-fold crystal
axis) in the continuum theory for crystals from the classes $C_{nv}$. The very existence of strong enough frustrating DMIs will tend to suppress
classical 3D long-range ordering and exacerbate the impact of quantum-fluctuations in a low-dimensional spin-system.\\

\noindent{\bf Experimental details NMR}

\noindent{\it Experimental determination of the hyperfine coupling}: Fig.~\ref{Kvschi} shows the Knight shift $K$ determined from the peak of the
spectra in 3~T and 15~T versus the macroscopic susceptibility $\chi$ of the powder sample in 3~T. Linear fits give hyperfine coupling constants of
$A_{hyp} \approx -0.27$\,kOe/$\mu_B$, and an orbital shift of $K_{orb} = -0.005$\,\% for 3 T and $A_{hyp} \approx -0.24$\,kOe/$\mu_B$, and $K_{orb} =
-0.002$\,\% for 15~T. This is comparable to the value extracted in LiCuVO$_4$ for a field perpendicular to the chain direction, $A_{perp} =
-0.19$\,kOe/$\mu_B$ \cite{Kegler06}. Anisotropic hyperfine coupling as can be determined by measurements on single crystals (see, e.g.,
\cite{Smerald11,Nawa13}) cannot be extracted since only powder susceptibility data are available.

\noindent{\it{Calculation of the hyperfine tensor}}: The local field
of the Cu electron spins ${\bf S}_j$ at the nuclear spin ${\bf I}_i$ of the
Li ${\bf h}_i = \sum_j \hat{A}_{ij}\langle {\bf S}_j \rangle$ is transferred by the dipolar hyperfine coupling tensor
\begin{eqnarray}
\hat{A}_{i} = \sum_j \hat{A}_{ij} =
\begin{pmatrix} A^{aa}_i & A^{ab}_i & A^{ac}_i \\ A^{ba}_i & A^{bb}_i & A^{bc}_i \\ A^{ca}_i & A^{cb}_i & A^{cc}_i  \end{pmatrix}
\label{Ahyp}
\end{eqnarray}
where $a$, $b$ and $c$ denote the crystallographic axes. In the paramagnetic
state $h_i$ is proportional to $\chi$, since $\langle {\bf S} \rangle
\sim \chi {\bf H}$. Compared to the case of \Licuv\ where the tensor Eq.~(\ref{Ahyp}) was determined by Nawa \textit{et al.} \cite{Nawa13} by
measurements on single crystals, \LCSO\ is only available as powder samples. In addition, there are two different Li sites in \LCSO\
(Fig.~\ref{structure}). Li(1) is coupled most strongly to six Cu spins in three nearest neighboring chains, and Li(2) is coupled to four Cu spins in
two nearest neighboring chains. The dipolar hyperfine coupling tensors for
 both Li sites have been calculated by lattice sum over a radius of
160\,\AA\ and are given in units [kOe/$\mu_B$]:
\begin{eqnarray}
\hat{A}_{1} =
\begin{pmatrix} -0.21 & 0.13 & -0.0005 \\ 0.13 & 0.27 & -1.0 \\ -0.0005 & -1.0 & -0.061  \end{pmatrix}
\end{eqnarray}
\begin{eqnarray}
\hat{A}_{2} =
\begin{pmatrix} -0.29 & 0.014 & 0.015 \\ 0.014 & -0.18 & -1.02 \\ 0.015 & -1.02 & 0.474  \end{pmatrix}
\end{eqnarray}
For both Li sites, there are diagonal elements which agree with the
hyperfine coupling determined from the $K$ vs. $\chi$ plot in
Fig.\ref{Kvschi}. This means, the main peak in the spectrum contains
intensity of both Li sites. Note, that also the off-diagonal elements may
contribute to the main peak due to the powder averaging of a specific
angular dependence of $h_i$, and that a small transferred hyperfine coupling
may exist, too.

The off-diagonal elements can transfer longitudinal electron spin fluctuations
in the paramagnetic state to transverse fluctuating hyperfine
fields at the Li site (cf. Eq.~(\ref{T1eqn}) in the main text). This would
lead to spin lattice relaxation by longitudinal spin fluctuations. On the
other hand, if spin correlations are peaked for certain wave vectors
 \textbf{Q}$_0$ in the short-range ordered helical, SDW or nematic states, phase
factors $\Theta(\textbf{q})$ may lead to a cancellation of these off-diagonal
elements making $T_1^{-1}$ insensitive to longitudinal spin
fluctuations \cite{Nawa13}. However, such a sharp peak at \textbf{Q}$_0$
is expected only for the field in the chain direction and not perpendicular
to it due to the weak intra-chain coupling \cite{Nawa13}. Thus, in the
measurements on a powder sample of \LCSO\ where all orientations contribute to
the NMR signal one can reasonably expect that both longitudinal and transverse
spin fluctuations give rise to the spin lattice relaxation at all
temperatures.
\begin{figure}
\includegraphics[width=0.9\columnwidth]{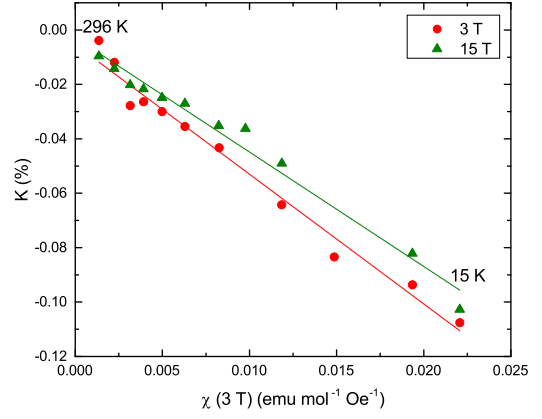}
\caption{Knight shift determined in 3 and 15\,T versus macroscopic susceptibility of the powder sample in 3\,T. The lines are fits as described in
the main text.} \label{Kvschi}
\end{figure}

\noindent{\it Spin lattice relaxation, fits and stretched exponential
relaxation}: Fig.\ref{T1fit} shows three exemplary fits of the nuclear
magnetization $M$ versus time $\tau$ for a magnetic field of 16~T at 20K,
4.2K, and 1.7 K. $M(\tau)$ could then be fitted to a single exponential
function $M(\tau) = M_0\{1-f\exp[-(\tau/T_1)^{b}]\}$ with $f=1$ for ideal
saturation. At 4.2 K, two fits are shown: one with a fixed stretching
exponent $b$ = 1, and one with $b$ as a free fitting parameter ($b$ = 0.81).
The difference is negligible and the extracted $T_1$ value is the same
within the error bars. Only at the lowest temperatures a stretching exponent
is necessary to fit the data, as can be seen for the data measured at
1.7\,K. The lower panel of Fig.~\ref{T1fit} shows the stretching exponent
$b$ for all fields and temperatures below 20\,K. Only at higher fields
$>13$\,T and very low $T$ a substantial distribution of spin lattice
relaxation rates appears.
\begin{figure}[h!]
\includegraphics[width=0.9\columnwidth]{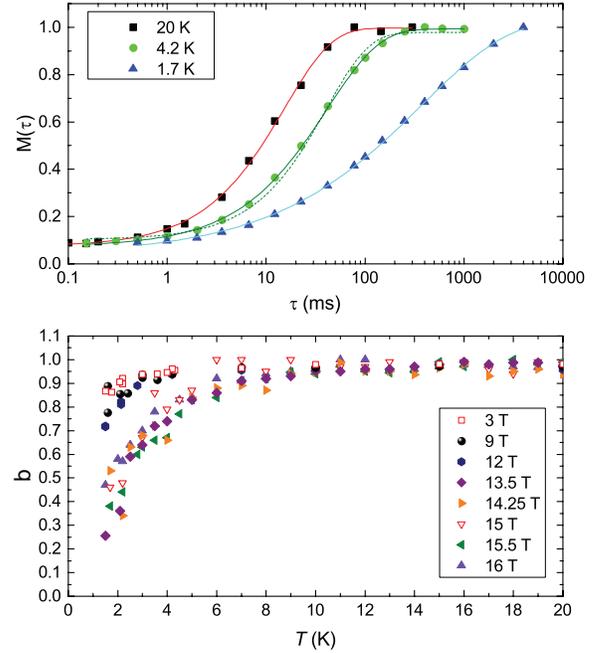}
\caption{Upper panel: Nuclear magnetization M versus $\tau$ in 16 T and for
three different temperatures. The lines are fits as described in the
text. The dashed line is a fit with $b$ fixed to 1. Lower panel: Stretching
exponent $b$ versus temperature for all different fields.} \label{T1fit}
\end{figure}
\begin{figure}[!h]
\includegraphics[width=0.9\columnwidth]{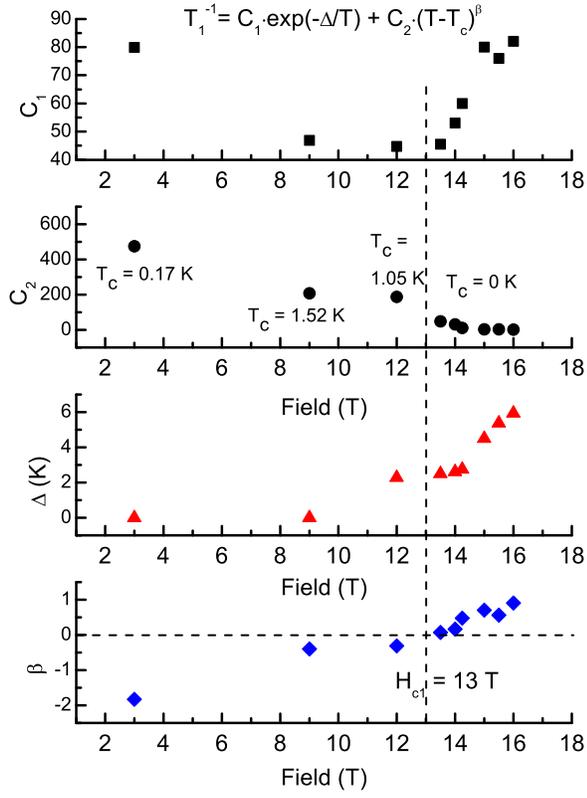}
\caption{Magnetic field dependence of the model parameters according to Eq.~(\ref{mainfit}). Vertical dashed line denotes the critical field $H_{c1}
= 13 $\,T corresponding to the isosbestic point in Fig.~\ref{Figisos}.} \label{T1fitparam}
\end{figure}
\noindent{\it Modelling of the \slrr(T) dependences}: Experimental \slrr(T) curves for all measured magnetic fields
(Fig.~\ref{T1}) were modelled with Eq.~(\ref{mainfit}). The fit parameters are presented in Fig.~\ref{T1fitparam}.

\clearpage


\end{document}